\documentclass[preprint]{JASA}
\usepackage{filecontents}

\usepackage{algpseudocode}
\usepackage[mode=buildnew]{standalone}
\usepackage{tikz}

\begin{document}

\title[Learnable STRFs]{Learning spectro-temporal representations of complex sounds with parameterized neural  networks}
\author{Rachid Riad}
\affiliation{Ecole des Hautes Etudes en Sciences Sociales, Centre National de la Recherche Scientifique, INRIA, Ecole Normale Supérieure-Paris Sciences \& Lettres University, 29 rue d’Ulm, 75005 Paris, France}
\affiliation{NeuroPsychologie Interventionnelle, Département d’Études Cognitives, Ecole Normale Supérieure, Institut national de la santé et de la recherche médicale, Institut Mondor de Recherche Biomédicale, Neuratris, Université Paris-Est Créteil, Paris Sciences \& Lettres University, 29 rue d’Ulm, 75005 Paris, France}

\author{Julien Karadayi}
\affiliation{Ecole des Hautes Etudes en Sciences Sociales, Centre National de la Recherche Scientifique, INRIA, Ecole Normale Supérieure-Paris Sciences \& Lettres University, 29 rue d’Ulm, 75005 Paris, France}

\author{Anne-Catherine Bachoud-Lévi}
\affiliation{NeuroPsychologie Interventionnelle, Département d’Études Cognitives, Ecole Normale Supérieure, Institut national de la santé et de la recherche médicale, Institut Mondor de Recherche Biomédicale, Neuratris, Université Paris-Est Créteil, Paris Sciences \& Lettres University, 29 rue d’Ulm, 75005 Paris, France}

\author{Emmanuel Dupoux}
\affiliation{Ecole des Hautes Etudes en Sciences Sociales, Centre National de la Recherche Scientifique, INRIA, Ecole Normale Supérieure-Paris Sciences \& Lettres University, 29 rue d’Ulm, 75005 Paris, France}
\affiliation{Facebook AI Research, Paris, France}


\date{\today}

\begin{abstract}



Deep Learning models have become potential candidates for auditory neuroscience research, thanks to their recent successes on a variety of auditory tasks. Yet, these models often lack interpretability to fully understand the exact computations that have been performed.
Here, we proposed a parametrized neural network layer, that computes specific spectro-temporal modulations based on Gabor kernels (Learnable STRFs) and that is fully interpretable. We evaluated predictive capabilities of this layer on Speech Activity Detection, Speaker Verification, Urban Sound Classification and Zebra Finch Call Type Classification. We found out that models based on Learnable STRFs are on par for all tasks with different toplines, and obtain the best performance for Speech Activity Detection. As this layer is fully interpretable, we used quantitative measures to describe the distribution of the learned spectro-temporal modulations.  The filters adapted to each task and focused mostly on low temporal and spectral modulations. The analyses show that the filters learned on human speech have similar spectro-temporal parameters as the ones measured directly in the human auditory cortex. Finally, we observed that the tasks organized in a meaningful way: the human vocalizations tasks closer to each other and bird vocalizations far away from human vocalizations and urban sounds tasks.

\end{abstract}


\maketitle


\section{\label{sec:1} Introduction}

The main objective of auditory neuroscience is to build models that can both predict the brain neural responses to relevant sounds and the behaviours associated with these responses \citep{kell2019deep,pillow2019editorial}. While most of the auditory neuroscience research has focused on the neural side, there is growing recognition for the importance to also match the performance of living organisms on a variety of behavioral tasks \citep{yarkoni2017choosing}. In recent years, major progress has been achieved with Deep Neural Networks (DNNs) which, after training with supervised classification objectives on large datasets, proved able to perform near human performance on a variety of audio tasks such as automatic speech recognition \citep{amodei2016deep}, speaker verification \citep{snyder2018x} or audio scene classification \citep{salamon2017deep}. These trained systems therefore become potential candidate models for auditory neuroscience \citep{koumura2019cascaded}, and have already started to be used to account for perceptual results \citep{saddler2020deep} and brain data \citep{kell2018task} in humans.

DNNs models typically take as input a spectral representation (although some new trends consist in side-stepping this representation and work directly from the raw waveform). Working from a spectral representation has biological plausibility, since it matches approximately what we know about the first stage of auditory processing \citep{stevens1937scale}. 
However, DNNs models are less biologically motivated regarding the next steps. Most of them use rather generic connectivity patterns (fully connected, convolutional or recurrent networks), which while being very powerful in learning task-specific representations from an engineering point of view, lack both interpretability and support in the auditory neuroscience.
To push the understanding of both the artificial and real neural networks, there have been some attempts to decode the representation extracted from biological measurements or computed by deep learning models \citep{thoret2020probing, ondel2019deriving}. Even though these methods allow to uncover the important aspects of the stimuli, they rely on simplifying hypotheses (linearity of the responses, independence across neurons \citep{meyer2017models, shamma1996auditory}) and they do not provide in depth explanation on how the DNNs made their decisions.



Fortunately, the stages beyond the extraction of the acoustic spectrum have been studied over the past few years with novel understanding of the representations and processing involved \citep{mcdermott2018audition}.
Slow spectral and temporal modulation built on top of the spectrum have been shown in psychophysical tests to be useful for several audio tasks solved by mammals: they contribute to speech intelligibility \citep{elliott2009modulation,elhilali2003spectro, edraki2019improvement}, they help to boost performance for speech processing in noisy environments \citep{mesgarani2006discrimination, chang2014robust, Vuong2020}. In addition, the responses to such spectral and temporal modulations of natural sounds can be decoded from human fMRI \citep{santoro2017reconstructing} and have been measured directly with invasive techniques in ferrets \citep{depireux2001spectro}, in birds \citep{woolley2005tuning}, and also in the human brain \citep{hullett2016human}.

Analytic models of these modulations have been proposed \citet{chi2005multiresolution, chang2014robust, schadler2012spectro, ezzat2007spectro}, on the basis of wavelet analysis. The idea is that on top of the spectrum, spectro-temporal wavelets or gabor patches can be defined which drive both behavioral responses and brain signals. The problem of such analytic models is that they only propose a potentially very large representation space, but provide no method to select which gabor patch is relevant for which task. But analyses of brain signals show that the responses from the auditory cortex are not fixed, but vary depending on the task at hand \citep{francis2018laminar, jaaskelainen2007short,fritz2003rapid}. Therefore, what is needed, is a model that can learn the characteristics of the spectro-temporal representations that are relevant to the task.

\begin{figure}[ht]
\includegraphics[width=0.8\reprintcolumnwidth, angle=0.0]{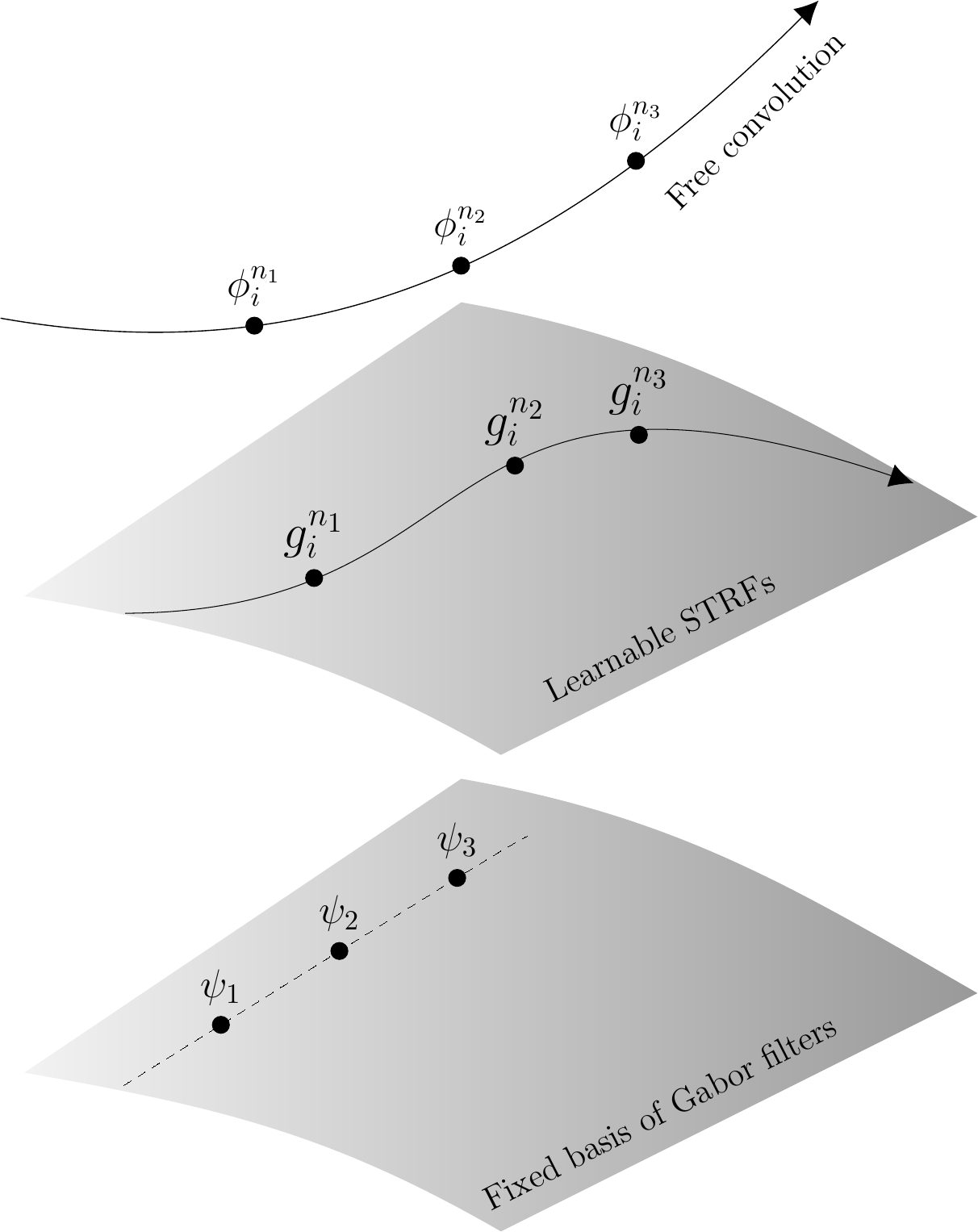}
\caption{\label{fig:Fig0}{Schematic illustration of the space of functions and different approaches to obtain spectro-temporal representations of sounds. (top) Free Learnable Convolution $(\phi^{n_i}_k)$ \citep{mlynarski2018learning, ondel2019deriving}; (middle) Learnable STRFs (Learnable spectro-temporal filters) $(g_k^{n_i})$ (this study); (bottom) Fixed basis of Gabor filters $(\psi_k)$ \citep{mesgarani2006discrimination, chang2014robust,bellur2015detection, schadler2012spectro, elie2016vocal}. The upper index $n_i$ and lower index $k_{th}$ represent the $n_i$-step during learning for the $k_{th}$ filter.}}

\raggedright
\vspace{-2.7em}
\end{figure}

This is the goal of this work. We introduce a parametrized neural network which explicitly represent spectro-temporal filters, but which parameters are differentiable and can therefore be tuned to each particular task. There are two advantages of this approach as illustrated in Figure \ref{fig:Fig0}. First, as Analytic Models, and contrary to standard DNN models, this model is fully interpretable. The parameters of each filter can be directly read off the model and compared to physiological or neural data. Second, as DNNs, but contrary to Analytic Models, this model can be tuned to different tasks, accounting both for behavioral results and for the task-specificity of the brain representations. As a side issue, since the model is constrained and has few parameters, it has the potential to explain perceptual learning aspect of plasticity with a lot less training data than is typically used in generic DNNs. Therefore, the model makes direct and testable predictions about the auditory representation as a function of the task.

The paper is organized as follows: Section~\ref{sec:methods} presents the Methods with our parametrized neural network model, and the different ways to analyze the distribution of the learned spectro-temporal modulations; in Section~\ref{sec:setup} we described the experimental setup with the different computational tasks, data, toplines, and evaluations; Section~\ref{sec:results} presents the performance results, the analysis of the learned distribution of spectro-temporal modulation for each setup and the discussions. Section~\ref{sec:conclusions} presents our conclusions and the potential future work.




\begin{figure}[ht]
\includegraphics[width=\reprintcolumnwidth, angle=0.0]{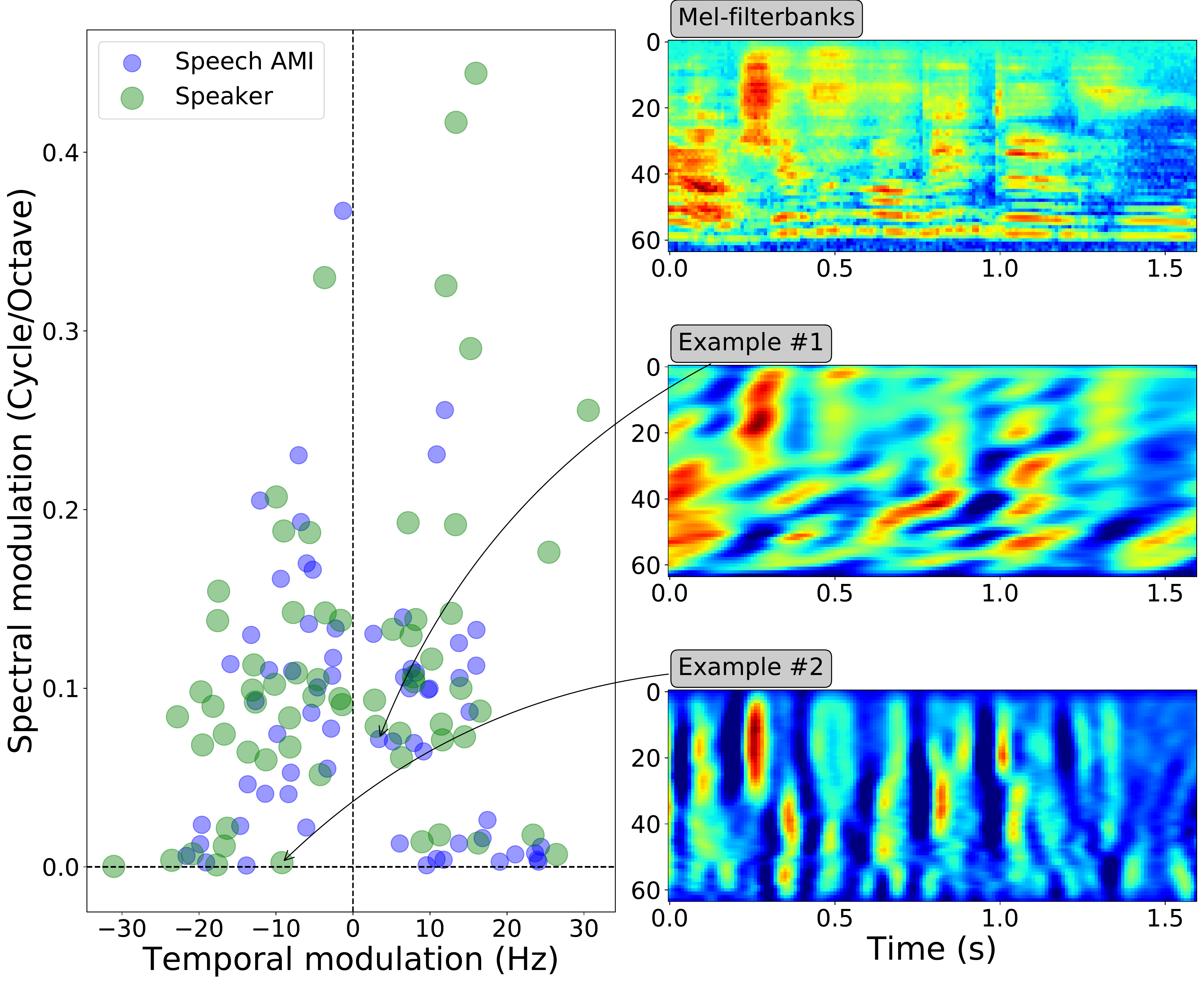}
\caption{\label{fig:ami_chime5_strf_omega}{Temporal and Spectral Modulations populations learned to tackle Speech Activity Detection (Speech AMI) on the AMI dataset, and Speaker Verification (Speaker) (left). Mel-filterbanks representation of a sentence pronounced by a Female Speaker (top right). Outputs examples computed by the convolution of specific learned STRF kernels with the input Mel-filterbanks (middle and bottom right).}}

\raggedright
\vspace{-2em}
\end{figure}




\section{Models and Methods}
\label{sec:methods}
\subsection{Learnable spectro-temporal filters\label{subsec:3.1}}
Here, $\Re$, $\Im$, $|.|$, $[.,.]$ represent the real part, imaginary part, modulus and the concatenation operators respectively. $\{.\}$ represents a set and $|{.}|$ the cardinal of a specific set.
\subsubsection{First stage of processing}
The first audio processing step is the transformation of the audio signal from the time domain into the frequency domain $\mathbf{Y}(t, f)$. Each excerpt of sound given to the network is normalized per instance with a 1 dimensional Instance Normalization layer \citet{ulyanov2016instance}. Then the sound is decomposed into a Filter banks spaced based on the Mel scale after a log compression (Mel-filterbanks) similarly as the resolution of the human cochlea. There are 64 filters with center frequencies in the range $[0.0, 8000.0 Hz]$. The computations for the Mel-filterbanks of the audio can be performed \textit{on-the-fly} directly on GPU thanks to \citet{nnaudio}.
\subsubsection{Definition of the Learnable spectro-temporal filters}

The second step of front-end processing is a set of convolution between the time-frequency representation of the audio and a set of Gabor Filters \citep{gabor1946theory}.


The 2-D Gabor filter kernel $g_k$ is a sine-wave $w_k$ modulated by a 2-D Gaussian envelope $s_k$.  Each Gabor filter $g_k$ is expressed based on the set of parameters  $( \sigma_{t},\sigma_{f}, \gamma_k, F_{k})$ in polar coordinates. We used the following formulation in this work:
\begin{subequations}
\begin{eqnarray}
g_{k}(t, f)&=& s_{k}(t, f) \cdot w_k(t,f)\\
s_{k}(t, f) &=&
\frac{1}{2 \pi \sigma_{t_k} \sigma_{f_k}} e^{-\frac{1}{2}\left(\frac{t^{2}}{\sigma_{t_k}^{2}}+\frac{f^{2}}{\sigma_{f_k}^{2}}\right)}\\
w_k(t,f) &=& e^{ j\left(2 \pi \left(F_{k} R_{\gamma_k}\right)  \right)} \\
       R_{\gamma_k} &=& t \cos(\gamma_k) + f \sin(\gamma_k)
\end{eqnarray}
\end{subequations}

We obtain a bank of $N$ filters $\{g_k(t,f)\}_{k=0..N-1}$. These bank of filters is convoluted with the time-frequency representation $\mathbf{Y}$ to obtain the 3D representation $\mathbf{Z}$.
\begin{equation}
    \mathbf{Z}(t, f,k)=\sum_{u, v} \mathbf{Y}\left(u, v\right) g_{k}(t-u, f-v) \in \mathbb{C}
\end{equation}

These filters and their parameters can be used in 2D Convolution neural Networks \citep{gabornet2019} in different ways: (1) as an \textit{Initialisation} (Free 2D conv. Gabor Init.) of a 2D convolution neural network and the 2D grid is tuned completely by back-propagation (as in \citep{chang2014robust, schadler2012spectro, ezzat2007spectro}), (2) or finally as \textit{Learnable spectro-temporal Filters} (Learnable STRFs), so the gradient descent is only performed on the set of parameters $(  \sigma_{t_k},\sigma_{f_k},F_k,\gamma_k)$. Indeed, all the operators to derive the Gabor Filter based on $(  \sigma_{t_k},\sigma_{f_k},F_k,\gamma_k)$ are differentiable almost everywhere: $x \rightarrow e^x, x \rightarrow \cos(x), x \rightarrow \sin(x), x \rightarrow x^2, x \rightarrow \frac{1}{x}, x \rightarrow \text{constant} \cdot x$. The 2D grid instantiated by the Gabor Filter used the parameters $(  \sigma_{t_k},\sigma_{f_k},F_k,\gamma_k)$ in each cell, therefore, the gradients are summed over the 2D grid for each parameter. Each Learnable STRF filters were given 9 Mel-frequency spectral filters and 1.1s of context, thus yielding a size of 9x111 for each filter.

Finally, the output representation $\mathbf{Z}$ is in the complex domain $\mathbb{C}$. To be used by classic neural network architectures, we concatenated $[\Re(\mathbf{Z}), \Im(\mathbf{Z})]$ to obtain the representation to be fed to the rest of each network. We denoted by \textit{Learnable STRFs} this specific front-end in all result tables, and by  \textit{Learned STRFs} once we examined these representations.


\subsubsection{Descriptive quantifiers of the distribution\label{subsec:quantifiers}}

 We used quantitative measures to describe the structure of the distribution of the learned spectro-temporal Convolutions. This is in the same spirit as \citet{singh2003modulation} for the 2-D Modulation Power Spectrum of sound ensembles, and Modulation Power Spectrum of the spectral-temporal receptive field of auditory neurons in ferrets \citet{depireux2001spectro}. We converted the learned parameters in Cartesian coordinates \citep{schadler2012spectro} with the Temporal Modulation $\omega_{k}$ and Spectral Modulation $\Omega_{k}$: $(\sigma_{t},\sigma_{f}, \omega_{k}, \Omega_{k})$, where $\omega_k=  F_k \cos(\gamma_k)$ and $\Omega_k=  F_k \sin(\gamma_k)$. We took the same convention as \citet{singh2003modulation, chi2005multiresolution} for the up-sweep and down-sweep modulations, and represented only half the plan due to the symmetry.

 We adapted the measures of Separability, Asymmetry,  Low-pass coefficient and  Starriness coefficients with the interpretable parameters obtained for each Supervised Learning tasks. As the Learned spectro-temporal receptive fields (Learned STRFs) self-organized to solve each task, we examined each of this parameter for each task. Each $\alpha$ is estimated with the bootstrap re-sampling method \citep{efron1994introduction}  on the Learned STRFs (100 bootstraps).


\paragraph{Asymmetry}
The distribution of the learned STRFs can show asymmetry preferences. The distribution is considered asymmetric if there are preferences for either down-sweeps or up-sweeps Learned STRFs.

\begin{equation}
    \alpha_{\text{asymmetry}}=\frac{|\{g_k \text{ s.t. } \omega_k>0\}|}{|\{g_k\}|} = \frac{|\{g_k \text{ s.t. } \omega_k>0\}|}{N}
\end{equation}

If $\alpha_{\text{asymmetry}} \approx 0$, the distribution of STRFs filters $\{g_k(t,f)\}_{k=0..N-1}$ is considered symmetric. If $\alpha_{\text{asymmetry}} > 0$, there are more up-sweeps than down-sweeps.

\paragraph{Low Pass Coefficient and Starriness}

It has been observed in \citet{singh2003modulation}, that most energy in Modulation Power Spectrum was concentrated in low spectral and temporal modulations for natural sounds. In addition, the higher spectral and temporal modulations were not distributed uniformly but were mostly along the axes. We derived two coefficients to quantify these phenomena with the Learned STRFs:

\begin{equation}
    \alpha_{\text{low}} = \frac{|\{g_k \text{ s.t. } |\omega_k|< \Delta_t, \Omega_k< \Delta_f\}|}{N} =  \frac{N_{\text{low}}}{N}
\end{equation}
For the temporal modulation low limit we opt as \citet{singh2003modulation} for $\Delta_t=16Hz$. The spectral modulation low limit is set to $\Delta_f=0.08$ $Cycle/Octave$. These parameters were chosen deliberately low as most learned modulations were mostly concentrated in low temporal and spectral modulations.
The parameter $\alpha_{star}$ to measure "stariness" of the distribution is the relative measure of distribution  excluding regions with high joint temporal and spectral modulations and the regions with low joint temporal and spectral modulations.

\begin{equation}
    \alpha_{\text{star}}=\frac{N_{\Delta_t} + N_{\Delta_f} - 2 \times N_{\text{low}}}{N - N_{\text{low}}}
\end{equation}

The quantities $N_{\Delta_t} = |\{g_k \text{ s.t. } |\omega_k|< \Delta_t\}|$ and $N_{\Delta_f} = |\{g_k \text{ s.t. } \Omega_k< \Delta_f\}|$ are the regions near the axes.



\paragraph{Separability}

To obtain a separability measure from the learned STRFs, we approximated the 2D-distribution $\mathcal{P}(\omega,\Omega)$ of the filters with Kernel Density Estimation with Gaussian Filters. Then we evaluate if the normalized 2D-distribution $\mathcal{P}$ can be factorized into a product of two independent functions  $\mathcal{P}(\omega,\Omega) = G(\omega)\cdot F(\Omega)$. To quantify the separability, we calculated as \citet{singh2003modulation} the singular value decomposition of the $\mathcal{P}(\omega,\Omega)$ obtained from each task:

\vspace{-0.7em}
\begin{equation}
    \mathcal{P}\left(\omega, \Omega\right)=\sum_{i=1}^{n} \lambda_{i} g_{i}\left(\omega\right) \cdot h_{i}\left(\Omega\right), \lambda_{1}>\lambda_{2}>\cdots>\lambda_{n}
\end{equation}
\vspace{-0.7em}

Then, we computed the ratio of first singular value relative to the sum of all singular values.

\begin{equation}
    \alpha_{\text{sep}} = \frac{\lambda_{1}}{\sum_{i=1}^{n}\lambda_{i}}
\end{equation}

If $\alpha_{\text{sep}} \approx 1$, the distribution of the learned STRFs can be considered separable.

\subsubsection{Measuring distance between tasks based on the learned STRF filters and optimal transport \label{subsec:ot}}

The $\alpha$ measures provide some descriptors allowing some comparison between the learned distributions. However, they only look at one view and aspect of the learned distributions at a time. There is no clear way to measure the distances between each task based on the $\alpha$.
Besides, these $\alpha$ measures do not take into account the learned Gaussian envelope parameters $(\sigma_{t_k},\sigma_{f_k})$. Here, the goal is to obtain a quantitative metric able to compare the distributions obtained from each task. Usually, researchers fall back to the Mahalanobis distance or an approximation of the KL-divergence to compare observations of two sets of points. Yet, these metrics either make modelling assumptions about the data (approximation of the underlying density functions that generated the data), or it is impossible to compare set of points with different cardinals.


The non-parametric, natural and most powerful way to compare distributions is to use optimal transport distances \citep{peyre2019computational}. We compared the different tasks by comparing the learned STRFs using the regularized version Sinkhorn distance \citep{cuturi2013sinkhorn, flamary2017pot}. Especially, this regularized version of the optimal transport distance allow fast computation of distances and multiple assignments between points.

We made the choice to compare two individual learned STRFs with the Euclidean distance $||.||$ We normalized along each axis/parameter to not privilege for a specific parameter variability. Based on each task we tackled in this work, we obtained a distribution of normalized learned parameters $\{  \sigma_{t_k},\sigma_{f_k},\omega_k,\Omega_k\}_{task}$ with the size $n_{task}$ being the total number of filters used for this task.  Therefore, equipped with the Euclidean distance to compare the individual filters, we can obtain the cost matrix between two tasks $M_{(task_a,task_b)} \in \mathbb{R}^{n_{task_a}\times n_{task_b}}$. We did not privilege any learned STRFs to build the distribution, therefore we attributed equal weight to each individual filter $w_{task}=(1/n_{task}) 1_{n_{task}}$. This allows to compare the different task if we have several models due to cross validation (\textit{Urban} and \textit{Bird}) or less filters for a specific task (\textit{Bird}).  If we denote, by $<.,.>_{F}$ the norm of Froebenius between matrices, the regularized distance $d_{\lambda}$ between two tasks is defined as:

\vspace{-0.7em}
\begin{equation}
\begin{array}{r}
d_{\lambda}=\min_{P}<P, M>_{F}-\lambda \cdot h(P) \\
\qquad \begin{aligned}
\text{s.t. } P 1_{n_{task_a}} &=w_{task_a} \\
P^{T} 1_{n_{task_b}} &=w_{task_b} \\
P &\in \mathbb{R}^{n_{task_a}\times n_{task_b}}_{+} \\
h(P) &= -\sum_{i,j}P_{i,j}\log(P_{i,j})\\
\lambda&=10^{-3}

\end{aligned}
\end{array}
\end{equation}
\vspace{-0.7em}

Therefore, we were able to obtain a proxy on how close/far are two different tasks $task_a$ and $task_b$ to each other based on the Sinkhorn-distance $d_{\lambda}(\{  \sigma_{t_k},\sigma_{f_k},\omega_k,\Omega_k\}_{task_a}, \{  \sigma_{t_k},\sigma_{f_k},\omega_k,\Omega_k\}_{task_b})$. Based on the distances between all tasks, we built a hierarchical cluster tree and represent these distances with a dendogram (See Figure \ref{fig:clustering}).

\section{Experimental Setup}
\label{sec:setup}

We compared the Learnable STRFs layer with strong toplines that have been lately introduced to solve each task as well more classic baselines for each task. We tried to keep the systems with Learnable STRFs as close as possible from topline systems.


\vspace{-0.7em}
\subsection{Speech Activity Detection}
The goal of Speech Activity Detection is to segment a given stream audio into portions of Speech or Non-Speech. We choose this task, as it allows us to examine what are the exact spectro-temporal modulations that makes standout speech in a audio stream with silences and background noises \citep{mesgarani2006discrimination}.

We conduct experiments with 2 challenging datasets with different characteristics:

The AMI database \citep{mccowan2005ami} a meeting dataset in English recorded  with multiple microphones in 3 different rooms. There are 180 different speakers in the datasets. Here, we focus on the \textit{AMI.SpeakerDiarization.MixHeadset} protocol as we are working only single channel feature analysis. We denoted by \textit{Speech AMI} the experiments and the distribution of learned STRFs on this dataset and this task.
The CHiME5 database \citep{barker2018is} is a dataset recorded at home during parties. Here, we focus also on single channel feature analysis with the \textit{CHiME5.SpeakerDiarization.U01} protocol. We denoted by \textit{Speech CHIME5} the experiments and the distribution of learned STRFs on this dataset and this task.


We compared different input front-end to tackle this task. We evaluated the Learnable STRFs (64 filters) with a contraction layer (CL) as well the Free 2D convolution with a contraction layer (CL). The contraction layer is a convolution layer taking the outputs at each time step of the Learnable STRFs to reduce the number of dimensions of $\mathbf{Z}$.
We compared these techniques with classic signal processing baselines used in speech processing: Mel-filterbanks (64 filters) and MFCC (19 coefficients, with their deltas and their delta-deltas). We also compared with the more recent parametrized neural network SincNet. SincNet is composed of parametrized sinc functions, which implement 80 band-pass filters (to replace directly more classic input spectral representations), and 3 temporal convolution/pooling layers.
All the input front-end are then fed to a stack of two layers of BiLSTM layers of dimension 128 and two forward layers of dimension 32 before a final decision layer.  The learning rate is controlled by a cyclical scheduler, each cycle lasting for 21 epochs. Data augmentation is performed directly on the waveform using additive noise based on the MUSAN database \citep{snyder2015musan} with a random target signal-to-noise ratio ranging from 5 to 20 dB. To evaluate Speech Activity detection, we used the Detection Error Rate (DetER):
\[\text{Detection Error Rate} =\frac{ { T_{\text{false alarm}} +  T_{\text{missed detection} }}}{T_{\text{total speech}}}\]
We also reported the Missed detection Rate(\%) and False Alarm Rate (\%). We used the implementation of the metrics from {\fontfamily{qcr}\selectfont pyannote.metrics}  \citep{pyannote.metrics} and all experiments were run with {\fontfamily{qcr}\selectfont pyannote.audio}  \citet{bredin2020pyannote}.

We ran an additional analysis  for the Speech Activity Detection Task  to compare the use of $\Re(\mathbf{Z})$, $\Im(\mathbf{Z})$, $|\mathbf{Z}|$ and $[\Re(\mathbf{Z}), \Im(\mathbf{Z})]$ (See Table \ref{tab:appendix_vad_results} in Appendix \ref{sec:appendix}).
\vspace{-0.7em}
\subsection{Speaker Verification}
The goal of the Speaker Verification task in speech processing is to accept or reject the hypothesis that a given speaker pronounced a given sentence. To do so, we learned an embedding function of any speech sequence of variable length. We examined this task, as it is believed that spectro-temporal modulations encode specifically the speaker information \citep{lei2012spectro, elliott2009modulation}.

We followed the same procedure as \citet{coria2020metric} to conduct experiments with the two version of the VoxCeleb databases: VoxCeleb2 \citep{chung2018voxceleb2} is used for training, and VoxCeleb1 \citep{nagrani2017voxceleb} is split into two parts for a development and test sets.
We compared two different input front-end for the speaker verification task. We compared the Learnable STRFs (64 filters) with a contraction layer (CL) and the SincNet front-end, as described in the Speech Activity Detection setup. Each model is trained with the Additive Angular Margin Loss ($\alpha=10, m=0.05$) with stochastic gradient descent with a learning rate of $0.01$.
We compared the different Speaker Verification approaches with the Equal Error Rate (EER). We also measured the performance of each approach when using the the S-normalization. We also reported the baseline performance of the I-vector system trained on VoxCeleb1 combined with Probabilistic Linear Discriminant Analysis (PLDA) \citep{dehak2010front}. We denoted by \textit{Speaker} the experiments and the distribution of learned STRFs on this dataset and this task.

\subsection{Urban Sound Classification}
The problem of Urban Sound Classification is to classify short excerpt of audio sounds into broad categories  (Ex: Car horns, Air Conditioner, Drilling).
We investigated the use of the Learnable STRFs for Urban sound classification, especially to test the use of spectro-temporal modulations for other type of sounds than animal (human or bird) vocalizations \citep{mlynarski2018learning}.

We followed the same evaluation procedure as \citet{salamon2017deep} to evaluate the experiments with the UrbanSound8K database \citep{salamon2014dataset}. The dataset is composed of 8732 excerpts of urban sounds from 10 categories (air conditioner, car horn, children playing, dog bark, drilling, engine idling, gun shot, jackhammer, siren, street music), and split in 10 separate folds. To compare with previous approaches, each model is evaluated by cross-validation on the 10 folds. We reported the Mean, Min and Max of the Accuracy across the 10 folds. We used the code-base from \citet{arnault2020urban} for the training and evaluations of the 2 approaches. The topline approach is the use of a Mel-filterbanks with the CNN10 architecture from \citet{kong2020panns}. For the Learnable STRFs approach, the first convolution layer of the CNN10 architecture (Free 2D convolution with size 3x3 with 64 filters) is replaced by the Learnable STRFs layer (64 filters) on top of the Mel-filterbanks described in section \ref{sec:methods}.
The models are trained with the RAdam optimizer \citep{liu2019variance} with LookAhead \citep{zhang2019lookahead}. We also reported the results from \citet{salamon2017deep} as baseline. We denoted by \textit{Urban} the experiments and the distribution of learned STRFs on this dataset and this task.

\subsection{Zebra Finch Call Type classification}


Finally, we examined the Zebra Finch Call Type classification task \citep{elie2016vocal}. The goal of this task is to classify short-excerpt of sounds into Call Type categories for the Zebra Finch bird. Indeed, it has been found by \citet{elie2016vocal} that several properties the acoustic space allow to separate to some extent the Call Types in the repertoire of Zebra Finches. We tried to stay as close as possible from the experimental protocol of \citet{elie2016vocal}. The dataset is composed of 3433 excerpts of Zebra Finches' calls from 11 categories ('Wsst or aggressive call','Begging calls', 'Distance call', 'Distress call','Long Tonal call', 'Nest call', 'Song','Tet call', 'Thuk call','Tuck call', 'Whine call') produced by Adults and Chicks. The calls were segmented to keep only the 3 first seconds of each excerpt and if the file is too short, the sound was zero-padded.

Each set of features and model was evaluated with a random cross-validation procedure, that took into account the nested format the database. $80\%$ of the birds were kept for training and $20\%$ for testing. 50 different permutations of excluded birds were obtained to generate 50 training and validation data sets. To compare the approaches, we computed the Mean, Min, Max of the Accuracy over the permutations.

We ran 4 different baselines for this task based on 2 different input features and 2 types of classifiers. We extracted the features introduced by \citet{elie2016vocal}: Predefined Acoustical Features (PAF) and the Modulation Power Spectrum (MPS). The PAF features are composed of 23 parameters extracted from Spectral envelope, Temporal Envelope and the Fundamental Frequency. (Mean, Min, Max, Std of the F0; Mean of F1; Mean of F2; Mean of F3; Saliency; RMS energy; Max of the Amplitude; Mean, Std, Skewness, Kurtosis, Entropy, first, second and third quartiles of the frequency power spectrum; Mean, Std, Skewness, Kurtosis, Entropy of the temporal envelope) The MPS representation is the amplitude spectrum of the 2D Fourier Transform applied on the spectrum representation of the sound waveform. The MPS extracts the spectro-temporal modulations in a fine-grained fashion and sum the contribution along the frequency axis. We tested both these input features with Linear Discriminant Analysis (LDA) and Random Forest (RF) classifiers as in \citet{elie2016vocal}.

Finally, we evaluated the potential of the Learnable STRFs (24 filters) for this task. We combined the Learnable STRFs with a simple linear layer to output directly the decision layer. The models were trained with the Adam optimizer \citep{kingma2014adam}.
We denote by \textit{Bird} the experiments and the distribution of learned STRFs on this dataset and this task.


\section{Results and Discussions}
\label{sec:results}
First, we analyzed the quantitative performances to perform the tasks for the Learnable STRFs for the different audio benchmarks. Then, we examined and compared, qualitatively and quantitatively, the statistics of the Learned STRFs representations.

\subsection{Quantitative performance on audio benchmarks \label{sec:eng_results}}

Overall, the performances of the Learnable STRFs is on par for all tasks with the different baselines. There is no skip connection between the Mel-filterbanks and the rest of each neural network that has been considered. This means that these Learned STRFs are in someway useful to perform each task, as this layer act as a filter. A degradation of performance means that it might be not fully sufficient to use spectro-temporal modulations to perform this specific task.

The objective results for the \textit{Speech Activity Detection} task are shown for all models in Table \ref{tab:results_speech}. Overall, learnable front-end approaches with injected prior improved over the classic signal processing baselines, Mel-filterbanks and MFCCs. Yet, the approaches with Free 2D conv. were not capable to improve over the classic signal processing baselines and had the worst performance, even for the convolution initialized with Gabor filters.
The best-performing models for this task, were the ones trained with the Learnable STRFs, and outperformed all the baselines. It is improving over the State-of-the-Art model with SincNet on the AMI dataset, and matched the performance on the CHIME5 dataset. Therefore, adding prior for spectro-temporal modulations was beneficial for Speech Activity Detection.
The closest work to our knowledge around speech activity detection is \citet{Vuong2020}, where they derived a layer that learn the spectro temporal modulation especially for Voice Type Discrimination in an industrial environment. The main difference with our work, is that they relied on the expression of the discrete implementation of the Hilbert transform. They also reported that parametrized Neural Network were better than free convolutions. They also used a long receptive field along the time axis, and a small receptive field along the frequency axis.

\begin{table}[!ht]

\caption{Speech Activity Detection results for the different approaches described. Init. stands for Initialisation. CL stands for contraction layer, it is a convolution (conv.) layer reducing the size of the tensor dimension after the convolution (Free 2D conv. or Learnable STRFs) on the Mel-filterbanks. The Free 2D conv. had the same grid size as the Learnable STRFs (9x111). Each input front-end is then fed to a 2-layer BiLSTM and 2 feed-forward layers. The best scores for each metric overall are in \textbf{bold}. MD stands for Missed detection rate. FA stands for False Alarm rate. DetER stands for Detection Error Rate. For all metrics, lower is better. }
\vspace{-0.7em}
\label{tab:results_speech}
\begin{adjustbox}{width=0.48\textwidth}
\begin{tabular}{lcccp{0.01\linewidth}>{\centering\arraybackslash}ccc}
\hline
\hline
		Database & \multicolumn{3}{c}{AMI} & & \multicolumn{3}{c}{CHIME5 }\\
\cline{2-4} \cline{6-8}
Metric & DetER & MD & FA & & DetER & MD & FA \\
\hline
Input front-end &  &  &  & &  &  &  \\
Mel-filterbanks & 7.7 & 2.6 & 5.1 & & 24.1 & 2.8 & 21.3 \\
MFCC & 6.3 & 2.7 & 3.5 & & 19.6 & 1.6 & 18.0 \\
SincNet \cite{ravanelli2018speaker}  & 6.0 & \textbf{2.4} & 3.6 & & \textbf{19.2 }& 1.7 & 17.6 \\

Free 2D conv. Random Init. + CL & 8.0 & 3.0 & 5.0 & & 26.5 & 0.6 & 25.9 \\
Free 2D conv. Gabor Init. + CL  & 7.9 & 2.5 & 5.3 & & 26.4 & \textbf{0.2} & 26.1 \\
Learnable STRFs + CL & \textbf{5.8} & \textbf{2.4} & \textbf{3.4} & & \textbf{19.2} & 3.1 & \textbf{16.1} \\
\hline
\hline
\end{tabular}
\end{adjustbox}

\end{table}

\begin{table}[ht]
\caption{Speaker Verification results for the different approaches described.  CL stands for contraction layer, it is a convolution layer reducing the size of the tensor dimension after the convolution (Learnable STRFs). The X-vector \citet{snyder2018x} is used after each input front-end. We evaluated the performance of the Speaker Verification with and without S-normalization \citet{coria2020metric}. We also reported the baseline performance of the I-vector combined with Probabilistic Linear Discriminant Analysis (PLDA) \citep{dehak2010front}. The best scores for each metric overall are in \textbf{bold}. For the EER, lower is better.}
\label{tab:results_spk_verif}
\vspace{-0.7em}
\begin{tabular}{lcc}
\hline\hline
Metric &EER &EER w/ S-norm
\\
\hline
Baseline & & \\
I-vectors+PLDA \citep{dehak2010front}\footnotemark[1] & 8.8 & -- \\
\hline
Input front-end &  & \\
SincNet \citep{coria2020metric} & \textbf{3.9}  & \textbf{3.5} \\
Learnable STRFs + CL & 6.4  & 6.1\\
\hline\hline
\end{tabular}
\footnotetext[1]{This result is directly extracted from their paper and was not replicated for this study.}

\end{table}

The results for the Speaker Verification task are reported in Table \ref{tab:results_spk_verif}. We found out that the SincNet that was designed initially for Speaker Recognition \citep{ravanelli2018speaker} is getting better results than the Learnable STRFs + CL. The S-normalization improved both systems.  This result differ with previous results reported by \citet{lei2012spectro} that the spectro-temporal modulations were useful for speaker recognition. One difference, that could explain this discrepancy, is the use of Bayesian models after the different features (HMM-GMM). Indeed, the X-vector \citep{snyder2015musan} was designed based on the latest progresses of Deep Learning research to tackle the Speaker recognition task and were validated initially on spectral representations of the audio.
Our results suggest that spectro-temporal modulations are not fully sufficient to distinguish speakers.
Harmonic structure was found useful for the Speaker Verification and Recognition task \citet{imperl1997study}. One of the hypothesis is that the learning of the harmonic structure is more difficult with the outputs Learnable STRFs layer than directly with the Mel-filterbanks.

\begin{table}[!ht]
\caption{Urban Sound Classification results for the different approaches described. The best score for the mean Accuracy over the 10 folds overall is in \textbf{bold}. the CNN10 architecture from \citet{kong2020panns} is used after each input front-end. Higher is better.}
\label{tab:results_urban}

\vspace{-0.7em}

\begin{tabular}{lcc}
\hline\hline
Accuracy & Mean & [Min - Max]
\\ \hline
Baseline &  &\\
SB-CNN \citep{salamon2017deep} \footnotemark[1] & 79  \% & [71\%-85\%]  \\
\hline
Input front-end &  &\\
Free 2D conv. 3by3 \citep{kong2020panns} & \textbf{84\%} & [76\%-93\%] \\
Learnable STRFs & 82\%  & [74\%-90\%]\\


\hline\hline
\end{tabular}
\footnotetext[1]{This result is directly extracted from their paper and was not replicated for this study.}
\end{table}

The performances for the Urban Sound Classification task are reported in Table \ref{tab:results_urban}. The accuracy of the Learnable STRFs is above the baseline approach from \citet{salamon2017deep} and is on par (slightly below) with the CNN10 architecture using Mel-filterbanks \citet{kong2020panns}. It was studied before in \citet{espi2015exploiting}, that the use of different sizes of the Spectral representation was increasing the performance of deep learning models for acoustic Event Detection. This suggests that the varying sizes of the focus on the Mel-filterbanks representations boost the performances, both in time and frequency. In our case, the model learned to focus through the fitting of the $(\sigma_f,\sigma_t)$ parameters.

\begin{table}[ht]
\caption{Zebra Finch Call Type classification results for the different approaches. The best scores for each metric overall are in \textbf{bold}. PAF stands for Predefined Acoustical Features and MPS for the Modulation Power Spectrum. LDA stands for Linear Discriminant Analysis. RF stands for Random Forest. The Learnable STRFs input front-end is combined with a simple linear model to output directly the decisions. Higher is better.}
\label{tab:results_birds}
\vspace{-0.7em}

\begin{adjustbox}{width=0.48\textwidth}
\begin{tabular}{lcc}
\hline\hline
Accuracy & Mean & [Min - Max]
\\
Chance level  & 17\% & [6\%-23\%] \\
\hline
Features + Model  &  & \\

PAF \citet{elie2016vocal} + LDA & 57\% & [43\%-71\%] \\

PAF \citet{elie2016vocal} + RF & 59\% & [47\%-68\%] \\

MPS \citet{elie2016vocal} + LDA & 41\% & [23\%-53\%] \\

MPS \citet{elie2016vocal} + RF & \textbf{69\%} & [49\%-84\%] \\

\hline
Input front-end  &  & \\

 Learnable STRFs & 43\% & [23 \%-73\%] \\
%

\hline\hline
\end{tabular}
\end{adjustbox}
\end{table}

Finally, the results for the Zebra Finch Call Type classification are in Table \ref{tab:results_birds}. On one hand, the PAF features depended slightly on the model used after for classification (LDA 57\% to RF 58\%) while the MPS had the worst performance overall with linear model such as LDA while the MPS had the best performance overall when combined with the RF (going from 41 \% to 69\%). The Learnable STRFs models was decoded with a simple linear layer, so the closest baseline is the combination of the MPS with the LDA. The Learnable STRFs perform below the PAF features and the MPS with RF. The performance with combination of features in the MPS with RF suggest that the model with Learnable STRFs could benefit greatly from Adaptive Neural Trees \citet{tanno2019adaptive} to perform the task. In addition, this encourages the use co-occurrences or anti-occurrences of the spectro-temporal patterns in models as in  \citet{mlynarski2018learning,mlynarski2019ecological}, since the routing in RF imply to measure joint patterns in the feature space of the MPS.

\subsection{Description of the learned filters}
\label{sec:description_results}
\begin{figure*}[!ht]
\includegraphics[width=2.\reprintcolumnwidth, angle=0.0]{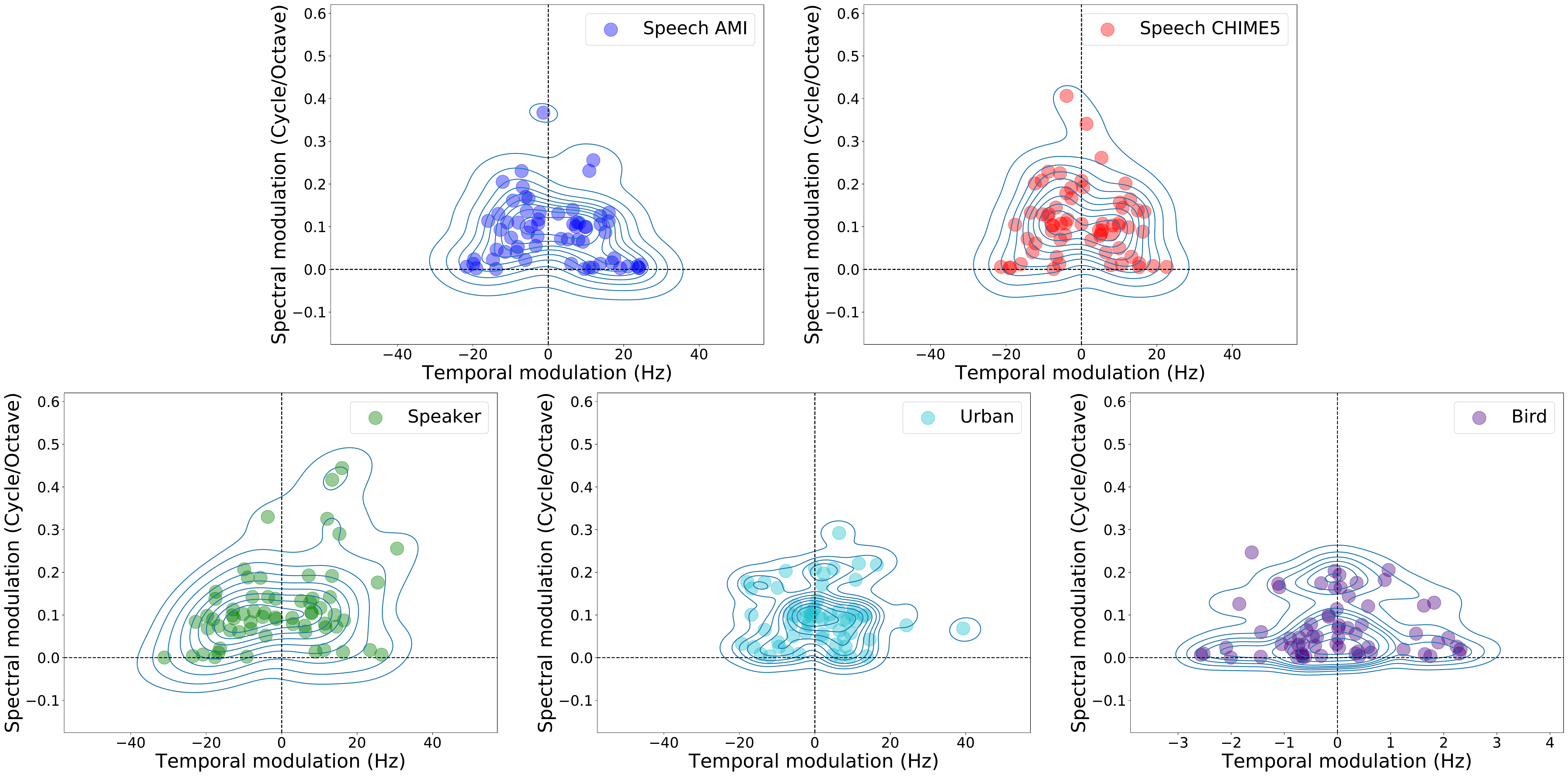}
\caption{\label{fig:all_strf_omega}{Temporal and Spectral Modulation of the Learned STRFs to tackle Speech Activity Detection on the AMI dataset (Speech AMI) and on the CHIME5 (Speech CHIME5), Speaker Verification on VoxCeleb (Speaker), Urban Sound Classification on Urban8k (Urban),  Zebra Finch Call Type  Classification (Bird). We displayed only a subset of the Learned STFRs of the Bird and Urban tasks for clarity. We also plotted the bi-variate distributions using kernel density estimation for each task.}}

\raggedright
\vspace{-1.1em}
\end{figure*}

\begin{figure*}[ht]
\includegraphics[width=2.\reprintcolumnwidth, angle=0.0]{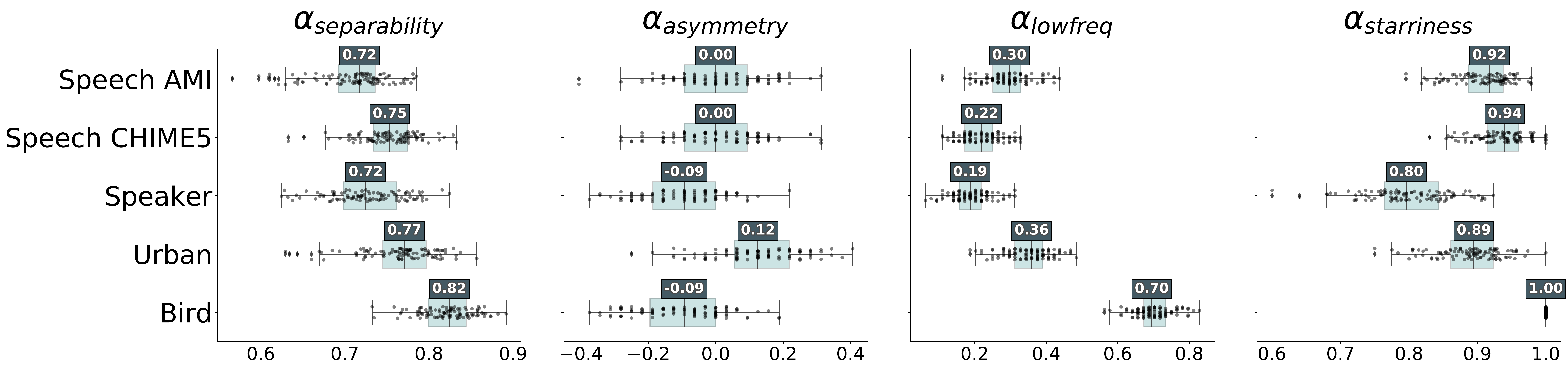}
\caption{\label{fig:alphas}{Separability,  asymmetry,  low-pass, starriness coefficients. Four quantifiers  that  measured different aspects learned distribution for the different tasks under study: Speech Activity Detection on the CHIME5 dataset (Speech CHIME5) and on the AMI (Speech AMI), Urban Sound Classification on Urban8k (Urban), Speaker Verification on VoxCeleb (Speaker) Zebra Finch Call Type  Classification (Bird). We displayed the median value for each $\alpha$ and task above each box-plot.}}

\vspace{-2.7em}

\raggedright

\end{figure*}

\begin{figure}[ht]
\includegraphics[width=\reprintcolumnwidth, angle=0.0]{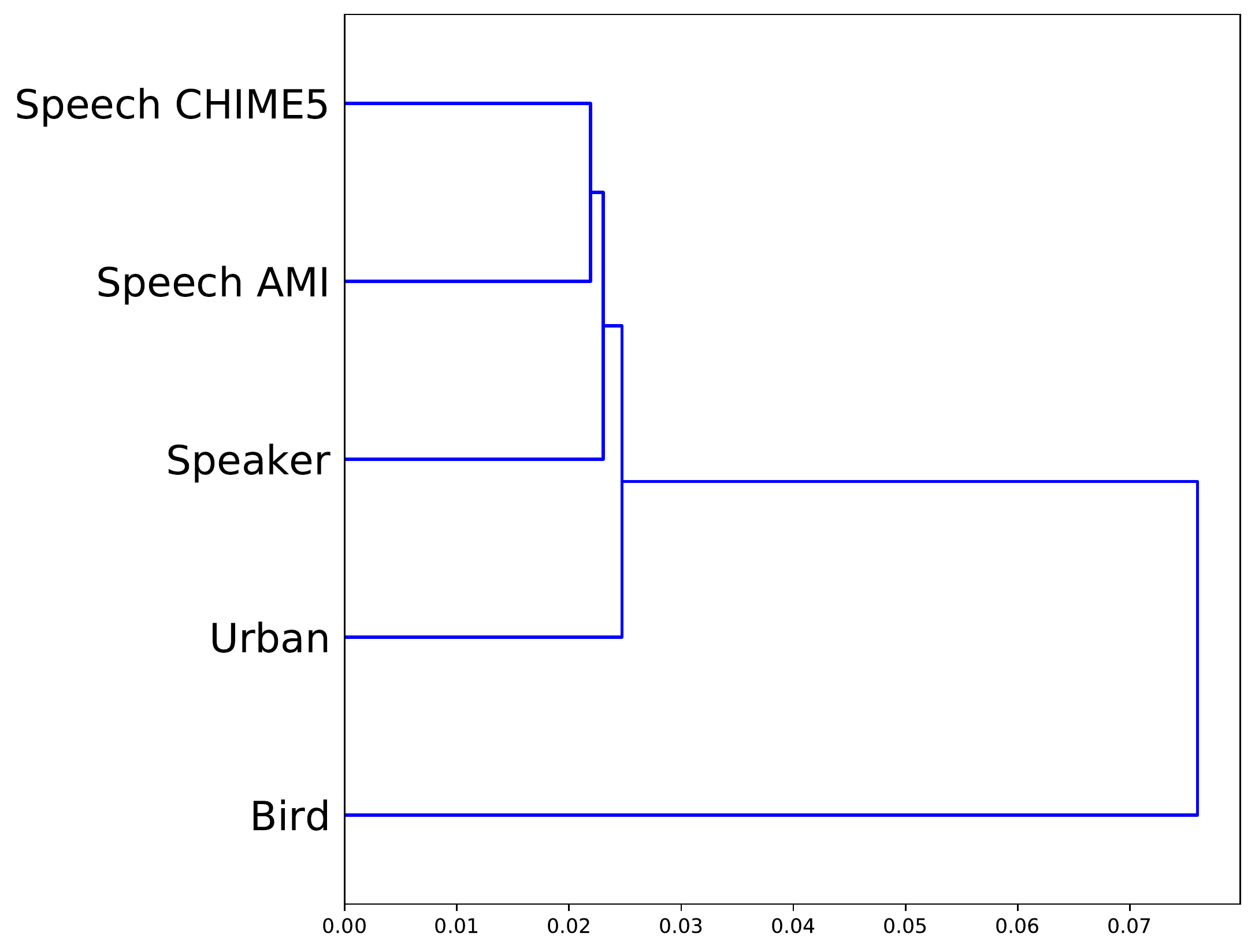}
\caption{\label{fig:clustering}{Hierarchical clustering of the tasks: Speech Activity Detection on the CHIME5 dataset (Speech CHIME5) and on the AMI (Speech AMI), Urban Sound Classification on Urban8k (Urban), Speaker Verification on VoxCeleb (Speaker) Zebra Finch Call Type Classification (Bird). The distance between tasks is computed between the learned STRF filters of each task with the Sinkhorn distance (we used the euclidean distance between each filter and the regularization parameter of the Sinkhorn distance is $\lambda=10^{-3}$). }}

\vspace{-1.3em}

\raggedright

\end{figure}

First, we observed that the Learned STRFs organized differently for each task, both the modulations $(\omega,\Omega)$ (See Figure \ref{fig:all_strf_omega}) and the size of the Gaussian envelopes through $(\sigma_t,\sigma_f)$ (See Figure \ref{fig:all_strf_sigmas} in Appendix).
Within the space allowed by the Nyquist theorem and the size of the convolutions, all the learned STRFs still concentrated in low spectral and temporal modulations (See Figure \ref{fig:all_strf_omega}). We also observed as \citet{singh2003modulation} that higher spectral modulations were found at low temporal modulations (and vice-versa). We found out that the Gaussian envelopes of the Learned STRFs can be characterized more by a continuum of values, not in a set of specific values. The Gaussian envelopes are more concentrated in the low values, and exhibit preferences  depending on the task for temporal or spectral shapes. Finally, the distributions of the Learned STRFs modulations and Gaussian envelopes of \textit{Speech} tasks on AMI and CHIME5 datasets, and the Speaker task look more similar than the \textit{Bird} and \textit{Urban} ones.
We quantified these observations with the description of the parameters described in Sub-section \ref{subsec:quantifiers} and the measure of distance between tasks with optimal transport in Sub-section \ref{subsec:ot}.

First, the separability index $\alpha_{separability}$ showed that most Learned STRFs are quite separable, and that the task related to human vocalizations (\textit{Speech} and \textit{Speaker}) were less separable than the other ones. We also found that all modulations have quite high $\alpha_{starriness}$ indexes. Similar results were found in \citet{singh2003modulation} for the separability and the starriness for the ensembles of sounds of Speech corpora, Zebra Finches vocalizations, and environmental sounds. \citet{schadler2012spectro} evaluated the use of high joint spectral and temporal modulation, and also found that they were degrading the performances for Speech Recognition tasks.

Besides, the Learned STRFs for the \textit{Speech} did not show preferences for up or down sweeps modulations ($\alpha_{asymmetry} \approx 0.0$), while the \textit{Speaker} and \textit{Bird} tasks exhibit slight preferences for down-sweeps and the \textit{Urban} for up-sweeps. The result for the \textit{Bird} task differed from \citet{singh2003modulation}. This could be explained by the fact that \citet{singh2003modulation} used a quantification of these parameters with an ensemble of sounds. The information about the specific characteristic of an individual Zebra Finch is mixed with the information of the Call Type. This suggests that a fully interpretable supervised approach might allow to decipher the different factors and contributions that influenced the acoustic properties of vocalizations.
Finally, the \textit{Bird} task focused more on the low frequency modulations ($\alpha_{lowfreq} \approx 0.70$) than the other tasks ($\alpha_{lowfreq} \leq 0.35$). We also observed that the Learned STRFs of the \textit{Speaker} task moved away from the low spectral modulations and yielded the lowest low-pass coefficient ($\alpha_{lowfreq} \approx 0.19$). Especially, \citet{elliott2009modulation} also found out that the removing of spectral modulations between 3 and 7 cycles/kHz significantly increases the gender mis-identifications of female speakers.
In addition, the results for the \textit{Speech} on the AMI and CHIME5 datasets and the \textit{Speaker} are very similar to the ones found directly in the auditory cortex neurons in awake monkeys \citep{massoudi2015spectrotemporal} and in awake humans \citep{hullett2016human}.
\citet{hullett2016human,massoudi2015spectrotemporal} measured responses of natural sounds directly in the superior temporal gyrus and found specific spectral modulation selectivity for $0.4\pm0.55$ Cycle/Octave and specific temporal modulation  $16\pm11$Hz; and most of the modulations were concentrated along the axes with high separability.


Finally, we examined the structure obtained from the hierarchical clustering based on the distances between tasks, see Sub-section \ref{subsec:ot} for the full description. We obtained the clustering tree in Figure \ref{fig:clustering}. We observed that the learned STRFs of the different tasks organized in a meaningful disposition. The Learned STRFs for \textit{Speech} on the CHIME5 and AMI are the closest to each other. Then, we found out that other Human vocalization task, \textit{Speaker}, is closer to the \textit{Speech} ones. On the other hand, the \textit{Bird} task organized far away from both the \textit{Urban} and the Human vocalizations tasks (\textit{Speech} and \textit{Speaker}). In future work, this method could be used to discover automatically phylogeny trees based only on acoustic properties of spectro-temporal modulations and test these predictions against a molecular-based phylogeny \citep{mccracken1997avian}.











\section{\label{sec:conclusions}Conclusion and future work}

In  summary, we examined the use of a parametrized neural network front-end to learn spectro-temporal modulations optimal for different behavioral tasks. This front-end, the Learnable STRFs, yielded performances close to published state-of-the-art using engineering-oriented neural network for Speaker Verification, Urban Sound Classification, Zebra Finch Call Type Classification, and obtained the best results on two datasets for Speech Activity Detection. As our front-end is fully interpretable, we found markedly different spectro-temporal modulations as a function of the task, showing that each task relies on a specific set of modulations. These task-specific modulations were globally congruent with previous work based on three approaches: spectro-temporal analysis of different audio signals \citep{elliott2009modulation}, analysis of trained neural networks \citep{schadler2012spectro}, and analysis of the auditory cortex \citep{santoro2017reconstructing,hullett2016human}. In particular, for the Speech Activity Detection task, we observed the same modulation distributions as the ones found directly the human auditory cortex listening while listening to naturalistic speech \citet{hullett2016human}. The modulations also displayed generic characteristics across tasks, namely, a predominance of low frequency spectral and temporal modulations and a high degree of 'stariness' and 'separability', corresponding to the fact that filters tend to remain close to either the temporal or spectral axis, with low occupation of joint spectro and temporal responses. This is consistent with \citet{singh2003modulation}. In order to encourage reproducible research, the developed Learnable STRFs layer, the learned STRFs modulations, and the recipes to replicate results are available in a open-source package \footnote{https://github.com/bootphon/learnable-strf}.


Several avenues of extensions are possible for this work, based on what is known in auditory neuroscience. First, this work only modelled the final outcome of plasticity after each task has been fully learned, starting from a random initialization. Yet, the same model could be used to address a range of issues relevant to changes occurring during task learning (top-down plasticity) or due to modification of the distribution of audio input (bottom-up plasticity). Recent work \citet{bellur2015detection} have investigated the adaptation of modulations in analytical models and witnessed several improvements in terms of engineering performances suggesting that this is also an interesting avenue in terms of behavioral modeling. Second, the analyses from \citet{hullett2016human} showed that not only neurons have spectro-temporal selectivity but are also topographically distributed along the posterior-to-anterior axis in the superior temporal gyrus. In future work, it would be interesting to reproduce such topography by using an auxiliary self-organizing maps objective in addition to the task-specific loss function for the STRFs. Finally, despite their wide use in auditory neuroscience, the spectro-temporal modulations do not provide a complete picture of computations in the auditory cortex \citep{williamson2016input}. A potential extension of our work would be to add an extra layer able to express co-occurrences and anti-occurrences of pairs of spectro-temporal receptive fields as in \citet{mlynarski2018learning,mlynarski2019ecological}. Such an extra layer would provide a learnable and interpretable extension to spectro-temporal representations.

To conclude, we emphasize that neuroscience-inspired parametrized neural networks can provide models that are both efficient in terms of behavioral tasks and interpretable in terms of auditory signal processing.



\vspace{-0.7em}

\begin{acknowledgments}
 This work is funded in part from the Agence Nationale pour la Recherche (ANR-17-EURE-0017 Frontcog, ANR-10-IDEX-0001-02 PSL*, ANR-19-P3IA-0001 PRAIRIE 3IA Institute). ACBL was funded through Neuratris, and ED in his EHESS role by Facebook AI Research (Research Gift) and CIFAR (Learning in Minds and Brains). The university (EHESS, CNRS, INRIA, ENS Paris-Sciences Lettres) obtained the datasets reported in this paper, and the experiment were run on it's computer resources.

\end{acknowledgments}

\vspace{-2.7em}
\appendix
\vspace{-0.7em}
\section{ \label{sec:appendix}}
\subsection{Importance of the representation of the Learnable STRFs}
\begin{figure*}[ht]
\includegraphics[width=1.8\reprintcolumnwidth, angle=0.0]{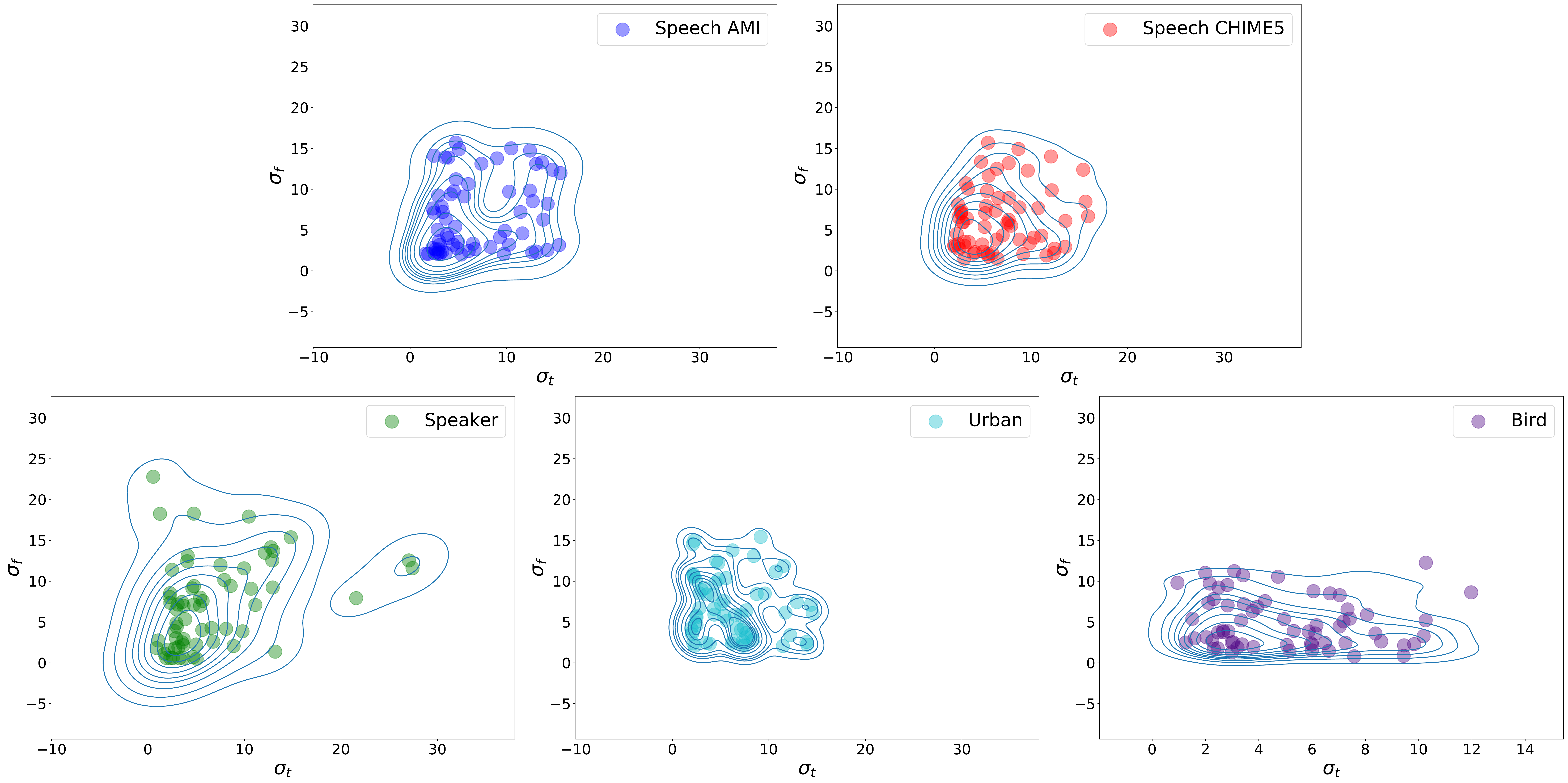}
\caption{\label{fig:all_strf_sigmas}{Gaussian envelopes $(\sigma_t, \sigma_f) $ of the Learned STRFs to tackle Speech Activity Detection on the AMI dataset (Speech AMI) and on the CHIME5 (Speech CHIME5), Speaker Verification on VoxCeleb (Speaker), Urban Sound Classification on Urban8k (Urban), Zebra Finch Call Type  Classification (Bird). We displayed only a subset of the Learned STFRs of the Bird and Urban tasks for clarity. We also plotted the bi-variate distributions using kernel density estimation for each task.}}
\vspace{-1.7em}

\raggedright

\end{figure*}






We performed an additional analysis of Speech Activity Detection of the choice of representations $\mathbf{Z}$ from the Learnable STRFs used in the subsequent neural network. The performance of the real part, the imaginary part and absolute values of the filter output are compared. The results are presented in Table \ref{tab:appendix_vad_results}. In comparison with the concatenation of the real and imaginary parts, the performance obtained for each part were in the same range on the AMI dataset, but were below on the CHIME5 dataset. As in \citet{schadler2012spectro}, this indicates that phase information contained in the real and imaginary parts is important for the Learnable STRFs.
\begin{table}[!ht]
\caption{Speech Activity Detection results for the different uses of the $\mathbf{Z}$ for the Learnable STRFs. CL stands for contraction layer, it is a convolution layer reducing the size of the tensor dimension after the convolution (Learnable STRFs). Each input front-end is then fed to a 2-layer BiLSTM and 2 feed-forward layers. The best scores for each metric overall are in \textbf{bold}. MD stands for Missed detection rate. FA stands for False Alarm rate. DetER stands for Detection Error Rate. For all metrics, lower is better.}
\vspace{-0.7em}

\begin{adjustbox}{width=0.48\textwidth}
\begin{tabular}{lcccp{0.01\linewidth}>{\centering\arraybackslash}ccc}
\hline
\hline
		Database & \multicolumn{3}{c}{AMI} & & \multicolumn{3}{c}{CHIME5 }\\
\cline{2-4} \cline{6-8}
Metric & DetER & MD & FA & & DetER & MD & FA \\
\hline
\multicolumn{3}{c}{Input front-end (Learnable STRFs + CL)}  &  & &  &  &  \\
Real part $\Re(\mathbf{Z})$ & 5.9 & 2.4 & 3.5 & &20.1 & 2.6 & 17.5 \\
Imaginary part $\Im(\mathbf{Z})$ & 5.9 & \textbf{2.2} & 3.7 & &22.1 & \textbf{1.0} & 21.1 \\
 Magnitude $|\mathbf{Z}|$ & 5.9 & \textbf{2.2} & 3.7& & 19.8 & 3.1 & 16.7 \\
Concatenation $[\Re(\mathbf{Z}), \Im(\mathbf{Z})]$ & \textbf{5.8} & 2.4 & \textbf{3.4} & & \textbf{19.2} & 3.1 & \textbf{16.1} \\
\hline
\hline
\end{tabular}
\end{adjustbox}
\vspace{-1.7em}
\label{tab:appendix_vad_results}
\end{table}






\begin{thebibliography}{67}
\def\enquote#1{``#1,''}
\def\plainquote#1{``#1''}
\expandafter\ifx\csname natexlab\endcsname\relax\def\natexlab#1{#1}\fi
\providecommand{\dourl}[1]{\href{http://#1}{\nolinkurl{#1}}}
\providecommand{\bibinfo}[2]{#2}
\providecommand{\noopsort}[1]{}
\providecommand{\switchargs}[2]{#2#1}
  \def\eatspace #1{#1}

\bibitem[{Alekseev and Bobe(2019)}]{gabornet2019}
\bibinfo{author}{Alekseev, A.},  and \bibinfo{author}{Bobe, A.}
  (\textbf{\bibinfo{year}{2019}}). \enquote{\bibinfo{title}{Gabornet: Gabor
  filters with learnable parameters in deep convolutional neural network}} pp.
  \bibinfo{pages}{1--4}, \dodoi{10.1109/EnT47717.2019.9030571}.

\bibitem[{Amodei \emph{et~al.}(2016)Amodei, Ananthanarayanan, Anubhai, Bai,
  Battenberg, Case, Casper, Catanzaro, Cheng, Chen
  \emph{et~al.}}]{amodei2016deep}
\bibinfo{author}{Amodei, D.}, \bibinfo{author}{Ananthanarayanan, S.},
  \bibinfo{author}{Anubhai, R.}, \bibinfo{author}{Bai, J.},
  \bibinfo{author}{Battenberg, E.}, \bibinfo{author}{Case, C.},
  \bibinfo{author}{Casper, J.}, \bibinfo{author}{Catanzaro, B.},
  \bibinfo{author}{Cheng, Q.}, \bibinfo{author}{Chen, G.} \emph{et~al.}
  (\textbf{\bibinfo{year}{2016}}). \enquote{\bibinfo{title}{Deep speech 2:
  End-to-end speech recognition in english and mandarin}} in
  \emph{\bibinfo{booktitle}{ICML}}, pp. \bibinfo{pages}{173--182}.

\bibitem[{Arnault \emph{et~al.}(2020)Arnault, Hanssens, and
  Riche}]{arnault2020urban}
\bibinfo{author}{Arnault, A.}, \bibinfo{author}{Hanssens, B.},  and
  \bibinfo{author}{Riche, N.} (\textbf{\bibinfo{year}{2020}}).
  \enquote{\bibinfo{title}{Urban sound classification: striving towards a fair
  comparison}} \bibinfo{journal}{arXiv preprint arXiv:2010.11805} .

\bibitem[{Barker \emph{et~al.}(2018)Barker, Watanabe, Vincent, and
  Trmal}]{barker2018is}
\bibinfo{author}{Barker, J.}, \bibinfo{author}{Watanabe, S.},
  \bibinfo{author}{Vincent, E.},  and \bibinfo{author}{Trmal, J.}
  (\textbf{\bibinfo{year}{2018}}). \enquote{\bibinfo{title}{The fifth `chime’
  speech separation and recognition challenge: Dataset, task and baselines}} in
  \emph{\bibinfo{booktitle}{Proceedings of the 19th Annual Conference of the
  International Speech Communication Association (INTERSPEECH 2018)}},
  \bibinfo{address}{Hyderabad, India}.

\bibitem[{Bellur and Elhilali(2015)}]{bellur2015detection}
\bibinfo{author}{Bellur, A.},  and \bibinfo{author}{Elhilali, M.}
  (\textbf{\bibinfo{year}{2015}}). \enquote{\bibinfo{title}{Detection of speech
  tokens in noise using adaptive spectrotemporal receptive fields}} in
  \emph{\bibinfo{booktitle}{2015 49th Annual Conference on Information Sciences
  and Systems (CISS)}}, \bibinfo{organization}{IEEE}, pp.
  \bibinfo{pages}{1--6}.

\bibitem[{Bredin(2017)}]{pyannote.metrics}
\bibinfo{author}{Bredin, H.} (\textbf{\bibinfo{year}{2017}}).
  \enquote{\bibinfo{title}{{pyannote.metrics: a toolkit for reproducible
  evaluation, diagnostic, and error analysis of speaker diarization systems}}}
  in \emph{\bibinfo{booktitle}{{Interspeech 2017, 18th Annual Conference of the
  International Speech Communication Association}}},
  \bibinfo{address}{Stockholm, Sweden},
  \dourl{http://pyannote.github.io/pyannote-metrics}.

\bibitem[{Bredin \emph{et~al.}(2020)Bredin, Yin, Coria, Gelly, Korshunov,
  Lavechin, Fustes, Titeux, Bouaziz, and Gill}]{bredin2020pyannote}
\bibinfo{author}{Bredin, H.}, \bibinfo{author}{Yin, R.},
  \bibinfo{author}{Coria, J.~M.}, \bibinfo{author}{Gelly, G.},
  \bibinfo{author}{Korshunov, P.}, \bibinfo{author}{Lavechin, M.},
  \bibinfo{author}{Fustes, D.}, \bibinfo{author}{Titeux, H.},
  \bibinfo{author}{Bouaziz, W.},  and \bibinfo{author}{Gill, M.-P.}
  (\textbf{\bibinfo{year}{2020}}). \enquote{\bibinfo{title}{Pyannote. audio:
  neural building blocks for speaker diarization}} in
  \emph{\bibinfo{booktitle}{ICASSP 2020-2020 IEEE International Conference on
  Acoustics, Speech and Signal Processing (ICASSP)}},
  \bibinfo{organization}{IEEE}, pp. \bibinfo{pages}{7124--7128}.

\bibitem[{Chang and Morgan(2014)}]{chang2014robust}
\bibinfo{author}{Chang, S.-Y.},  and \bibinfo{author}{Morgan, N.}
  (\textbf{\bibinfo{year}{2014}}). \enquote{\bibinfo{title}{Robust cnn-based
  speech recognition with gabor filter kernels}} in
  \emph{\bibinfo{booktitle}{Fifteenth annual conference of the international
  speech communication association}}.

\bibitem[{{Cheuk} \emph{et~al.}(2020){Cheuk}, {Anderson}, {Agres}, and
  {Herremans}}]{nnaudio}
\bibinfo{author}{{Cheuk}, K.~W.}, \bibinfo{author}{{Anderson}, H.},
  \bibinfo{author}{{Agres}, K.},  and \bibinfo{author}{{Herremans}, D.}
  (\textbf{\bibinfo{year}{2020}}). \enquote{\bibinfo{title}{nnaudio: An
  on-the-fly gpu audio to spectrogram conversion toolbox using 1d convolutional
  neural networks}} \bibinfo{journal}{IEEE Access} \textbf{In press}.

\bibitem[{Chi \emph{et~al.}(2005)Chi, Ru, and Shamma}]{chi2005multiresolution}
\bibinfo{author}{Chi, T.}, \bibinfo{author}{Ru, P.},  and
  \bibinfo{author}{Shamma, S.~A.} (\textbf{\bibinfo{year}{2005}}).
  \enquote{\bibinfo{title}{Multiresolution spectrotemporal analysis of complex
  sounds}} \bibinfo{journal}{The Journal of the Acoustical Society of America}
  \textbf{118}(2), \bibinfo{pages}{887--906}.

\bibitem[{Chung \emph{et~al.}(2018)Chung, Nagrani, and
  Zisserman}]{chung2018voxceleb2}
\bibinfo{author}{Chung, J.~S.}, \bibinfo{author}{Nagrani, A.},  and
  \bibinfo{author}{Zisserman, A.} (\textbf{\bibinfo{year}{2018}}).
  \enquote{\bibinfo{title}{Voxceleb2: Deep speaker recognition}}
  \bibinfo{journal}{Proc. Interspeech 2018} \bibinfo{pages}{1086--1090}.

\bibitem[{Coria \emph{et~al.}(2020)Coria, Bredin, Ghannay, and
  Rosset}]{coria2020metric}
\bibinfo{author}{Coria, J.~M.}, \bibinfo{author}{Bredin, H.},
  \bibinfo{author}{Ghannay, S.},  and \bibinfo{author}{Rosset, S.}
  (\textbf{\bibinfo{year}{2020}}). \enquote{\bibinfo{title}{{A Comparison of
  Metric Learning Loss Functions for End-To-End Speaker Verification}}} in
  \emph{\bibinfo{booktitle}{Statistical Language and Speech Processing}},
  edited by \bibinfo{editor}{L.~Espinosa-Anke},
  \bibinfo{editor}{C.~Mart{\'i}n-Vide}, and \bibinfo{editor}{I.~Spasi{\'{c}}},
  \bibinfo{publisher}{Springer International Publishing},
  \bibinfo{address}{Cham}, pp. \bibinfo{pages}{137--148}.

\bibitem[{Cuturi(2013)}]{cuturi2013sinkhorn}
\bibinfo{author}{Cuturi, M.} (\textbf{\bibinfo{year}{2013}}).
  \enquote{\bibinfo{title}{Sinkhorn distances: Lightspeed computation of
  optimal transport}} \bibinfo{journal}{Advances in neural information
  processing systems} \textbf{26}, \bibinfo{pages}{2292--2300}.

\bibitem[{Dehak \emph{et~al.}(2010)Dehak, Kenny, Dehak, Dumouchel, and
  Ouellet}]{dehak2010front}
\bibinfo{author}{Dehak, N.}, \bibinfo{author}{Kenny, P.~J.},
  \bibinfo{author}{Dehak, R.}, \bibinfo{author}{Dumouchel, P.},  and
  \bibinfo{author}{Ouellet, P.} (\textbf{\bibinfo{year}{2010}}).
  \enquote{\bibinfo{title}{Front-end factor analysis for speaker verification}}
  \bibinfo{journal}{IEEE Transactions on Audio, Speech, and Language
  Processing} \textbf{19}(4), \bibinfo{pages}{788--798}.

\bibitem[{Depireux \emph{et~al.}(2001)Depireux, Simon, Klein, and
  Shamma}]{depireux2001spectro}
\bibinfo{author}{Depireux, D.~A.}, \bibinfo{author}{Simon, J.~Z.},
  \bibinfo{author}{Klein, D.~J.},  and \bibinfo{author}{Shamma, S.~A.}
  (\textbf{\bibinfo{year}{2001}}). \enquote{\bibinfo{title}{Spectro-temporal
  response field characterization with dynamic ripples in ferret primary
  auditory cortex}} \bibinfo{journal}{Journal of neurophysiology}
  \textbf{85}(3), \bibinfo{pages}{1220--1234}.

\bibitem[{Edraki \emph{et~al.}(2019)Edraki, Chan, Jensen, and
  Fogerty}]{edraki2019improvement}
\bibinfo{author}{Edraki, A.}, \bibinfo{author}{Chan, W.-Y.},
  \bibinfo{author}{Jensen, J.},  and \bibinfo{author}{Fogerty, D.}
  (\textbf{\bibinfo{year}{2019}}). \enquote{\bibinfo{title}{Improvement and
  assessment of spectro-temporal modulation analysis for speech intelligibility
  estimation}} in \emph{\bibinfo{booktitle}{Interspeech 2019Annual Conference
  of the International Speech Communication Association}},
  \bibinfo{organization}{ISCA}, pp. \bibinfo{pages}{1378--1382}.

\bibitem[{Efron and Tibshirani(1994)}]{efron1994introduction}
\bibinfo{author}{Efron, B.},  and \bibinfo{author}{Tibshirani, R.~J.}
  (\textbf{\bibinfo{year}{1994}}). \emph{\bibinfo{title}{An introduction to the
  bootstrap}} (\bibinfo{publisher}{CRC press}).

\bibitem[{Elhilali \emph{et~al.}(2003)Elhilali, Chi, and
  Shamma}]{elhilali2003spectro}
\bibinfo{author}{Elhilali, M.}, \bibinfo{author}{Chi, T.},  and
  \bibinfo{author}{Shamma, S.~A.} (\textbf{\bibinfo{year}{2003}}).
  \enquote{\bibinfo{title}{A spectro-temporal modulation index (stmi) for
  assessment of speech intelligibility}} \bibinfo{journal}{Speech
  communication} \textbf{41}(2-3), \bibinfo{pages}{331--348}.

\bibitem[{Elie and Theunissen(2016)}]{elie2016vocal}
\bibinfo{author}{Elie, J.~E.},  and \bibinfo{author}{Theunissen, F.~E.}
  (\textbf{\bibinfo{year}{2016}}). \enquote{\bibinfo{title}{The vocal
  repertoire of the domesticated zebra finch: a data-driven approach to
  decipher the information-bearing acoustic features of communication signals}}
  \bibinfo{journal}{Animal cognition} \textbf{19}(2),
  \bibinfo{pages}{285--315}.

\bibitem[{Elliott and Theunissen(2009)}]{elliott2009modulation}
\bibinfo{author}{Elliott, T.~M.},  and \bibinfo{author}{Theunissen, F.~E.}
  (\textbf{\bibinfo{year}{2009}}). \enquote{\bibinfo{title}{The modulation
  transfer function for speech intelligibility}} \bibinfo{journal}{PLoS
  computational biology} \textbf{5}(3).

\bibitem[{Espi \emph{et~al.}(2015)Espi, Fujimoto, Kinoshita, and
  Nakatani}]{espi2015exploiting}
\bibinfo{author}{Espi, M.}, \bibinfo{author}{Fujimoto, M.},
  \bibinfo{author}{Kinoshita, K.},  and \bibinfo{author}{Nakatani, T.}
  (\textbf{\bibinfo{year}{2015}}). \enquote{\bibinfo{title}{Exploiting
  spectro-temporal locality in deep learning based acoustic event detection}}
  \bibinfo{journal}{EURASIP Journal on Audio, Speech, and Music Processing}
  \textbf{2015}(1), \bibinfo{pages}{1--12}.

\bibitem[{Ezzat \emph{et~al.}(2007)Ezzat, Bouvrie, and
  Poggio}]{ezzat2007spectro}
\bibinfo{author}{Ezzat, T.}, \bibinfo{author}{Bouvrie, J.},  and
  \bibinfo{author}{Poggio, T.} (\textbf{\bibinfo{year}{2007}}).
  \enquote{\bibinfo{title}{Spectro-temporal analysis of speech using 2-d gabor
  filters}} in \emph{\bibinfo{booktitle}{Eighth Annual Conference of the
  International Speech Communication Association}}.

\bibitem[{Flamary and Courty(2017)}]{flamary2017pot}
\bibinfo{author}{Flamary, R.},  and \bibinfo{author}{Courty, N.}
  (\textbf{\bibinfo{year}{2017}}). \plainquote{\bibinfo{title}{Pot python
  optimal transport library}} \dourl{https://pythonot.github.io/}.

\bibitem[{Francis \emph{et~al.}(2018)Francis, Elgueda, Englitz, Fritz, and
  Shamma}]{francis2018laminar}
\bibinfo{author}{Francis, N.~A.}, \bibinfo{author}{Elgueda, D.},
  \bibinfo{author}{Englitz, B.}, \bibinfo{author}{Fritz, J.~B.},  and
  \bibinfo{author}{Shamma, S.~A.} (\textbf{\bibinfo{year}{2018}}).
  \enquote{\bibinfo{title}{Laminar profile of task-related plasticity in ferret
  primary auditory cortex}} \bibinfo{journal}{Scientific reports}
  \textbf{8}(1), \bibinfo{pages}{1--10}.

\bibitem[{Fritz \emph{et~al.}(2003)Fritz, Shamma, Elhilali, and
  Klein}]{fritz2003rapid}
\bibinfo{author}{Fritz, J.}, \bibinfo{author}{Shamma, S.},
  \bibinfo{author}{Elhilali, M.},  and \bibinfo{author}{Klein, D.}
  (\textbf{\bibinfo{year}{2003}}). \enquote{\bibinfo{title}{Rapid task-related
  plasticity of spectrotemporal receptive fields in primary auditory cortex}}
  \bibinfo{journal}{Nature neuroscience} \textbf{6}(11),
  \bibinfo{pages}{1216--1223}.

\bibitem[{Gabor(1946)}]{gabor1946theory}
\bibinfo{author}{Gabor, D.} (\textbf{\bibinfo{year}{1946}}).
  \enquote{\bibinfo{title}{Theory of communication. part 1: The analysis of
  information}} \bibinfo{journal}{Journal of the Institution of Electrical
  Engineers-Part III: Radio and Communication Engineering} \textbf{93}(26),
  \bibinfo{pages}{429--441}.

\bibitem[{Hullett \emph{et~al.}(2016)Hullett, Hamilton, Mesgarani, Schreiner,
  and Chang}]{hullett2016human}
\bibinfo{author}{Hullett, P.~W.}, \bibinfo{author}{Hamilton, L.~S.},
  \bibinfo{author}{Mesgarani, N.}, \bibinfo{author}{Schreiner, C.~E.},  and
  \bibinfo{author}{Chang, E.~F.} (\textbf{\bibinfo{year}{2016}}).
  \enquote{\bibinfo{title}{Human superior temporal gyrus organization of
  spectrotemporal modulation tuning derived from speech stimuli}}
  \bibinfo{journal}{Journal of Neuroscience} \textbf{36}(6),
  \bibinfo{pages}{2014--2026}.

\bibitem[{Imperl \emph{et~al.}(1997)Imperl, Ka{\v{c}}i{\v{c}}, and
  Horvat}]{imperl1997study}
\bibinfo{author}{Imperl, B.}, \bibinfo{author}{Ka{\v{c}}i{\v{c}}, Z.},  and
  \bibinfo{author}{Horvat, B.} (\textbf{\bibinfo{year}{1997}}).
  \enquote{\bibinfo{title}{A study of harmonic features for the speaker
  recognition}} \bibinfo{journal}{Speech communication} \textbf{22}(4),
  \bibinfo{pages}{385--402}.

\bibitem[{J{\"a}{\"a}skel{\"a}inen \emph{et~al.}(2007)J{\"a}{\"a}skel{\"a}inen,
  Ahveninen, Belliveau, Raij, and Sams}]{jaaskelainen2007short}
\bibinfo{author}{J{\"a}{\"a}skel{\"a}inen, I.~P.}, \bibinfo{author}{Ahveninen,
  J.}, \bibinfo{author}{Belliveau, J.~W.}, \bibinfo{author}{Raij, T.},  and
  \bibinfo{author}{Sams, M.} (\textbf{\bibinfo{year}{2007}}).
  \enquote{\bibinfo{title}{Short-term plasticity in auditory cognition}}
  \bibinfo{journal}{Trends in neurosciences} \textbf{30}(12),
  \bibinfo{pages}{653--661}.

\bibitem[{Kell and McDermott(2019)}]{kell2019deep}
\bibinfo{author}{Kell, A.~J.},  and \bibinfo{author}{McDermott, J.~H.}
  (\textbf{\bibinfo{year}{2019}}). \enquote{\bibinfo{title}{Deep neural network
  models of sensory systems: windows onto the role of task constraints}}
  \bibinfo{journal}{Current opinion in neurobiology} \textbf{55},
  \bibinfo{pages}{121--132}.

\bibitem[{Kell \emph{et~al.}(2018)Kell, Yamins, Shook, Norman-Haignere, and
  McDermott}]{kell2018task}
\bibinfo{author}{Kell, A.~J.}, \bibinfo{author}{Yamins, D.~L.},
  \bibinfo{author}{Shook, E.~N.}, \bibinfo{author}{Norman-Haignere, S.~V.},
  and \bibinfo{author}{McDermott, J.~H.} (\textbf{\bibinfo{year}{2018}}).
  \enquote{\bibinfo{title}{A task-optimized neural network replicates human
  auditory behavior, predicts brain responses, and reveals a cortical
  processing hierarchy}} \bibinfo{journal}{Neuron} \textbf{98}(3),
  \bibinfo{pages}{630--644}.

\bibitem[{Kingma and Ba(2014)}]{kingma2014adam}
\bibinfo{author}{Kingma, D.~P.},  and \bibinfo{author}{Ba, J.}
  (\textbf{\bibinfo{year}{2014}}). \enquote{\bibinfo{title}{Adam: A method for
  stochastic optimization}} \bibinfo{journal}{arXiv preprint arXiv:1412.6980} .

\bibitem[{Kong \emph{et~al.}(2020)Kong, Cao, Iqbal, Wang, Wang, and
  Plumbley}]{kong2020panns}
\bibinfo{author}{Kong, Q.}, \bibinfo{author}{Cao, Y.}, \bibinfo{author}{Iqbal,
  T.}, \bibinfo{author}{Wang, Y.}, \bibinfo{author}{Wang, W.},  and
  \bibinfo{author}{Plumbley, M.~D.} (\textbf{\bibinfo{year}{2020}}).
  \enquote{\bibinfo{title}{Panns: Large-scale pretrained audio neural networks
  for audio pattern recognition}} \bibinfo{journal}{IEEE/ACM Transactions on
  Audio, Speech, and Language Processing} \textbf{28},
  \bibinfo{pages}{2880--2894}.

\bibitem[{Koumura \emph{et~al.}(2019)Koumura, Terashima, and
  Furukawa}]{koumura2019cascaded}
\bibinfo{author}{Koumura, T.}, \bibinfo{author}{Terashima, H.},  and
  \bibinfo{author}{Furukawa, S.} (\textbf{\bibinfo{year}{2019}}).
  \enquote{\bibinfo{title}{Cascaded tuning to amplitude modulation for natural
  sound recognition}} \bibinfo{journal}{Journal of Neuroscience}
  \textbf{39}(28), \bibinfo{pages}{5517--5533}.

\bibitem[{Lei \emph{et~al.}(2012)Lei, Meyer, and Mirghafori}]{lei2012spectro}
\bibinfo{author}{Lei, H.}, \bibinfo{author}{Meyer, B.~T.},  and
  \bibinfo{author}{Mirghafori, N.} (\textbf{\bibinfo{year}{2012}}).
  \enquote{\bibinfo{title}{Spectro-temporal gabor features for speaker
  recognition}} in \emph{\bibinfo{booktitle}{2012 IEEE International Conference
  on Acoustics, Speech and Signal Processing (ICASSP)}},
  \bibinfo{organization}{IEEE}, pp. \bibinfo{pages}{4241--4244}.

\bibitem[{Liu \emph{et~al.}(2019)Liu, Jiang, He, Chen, Liu, Gao, and
  Han}]{liu2019variance}
\bibinfo{author}{Liu, L.}, \bibinfo{author}{Jiang, H.}, \bibinfo{author}{He,
  P.}, \bibinfo{author}{Chen, W.}, \bibinfo{author}{Liu, X.},
  \bibinfo{author}{Gao, J.},  and \bibinfo{author}{Han, J.}
  (\textbf{\bibinfo{year}{2019}}). \enquote{\bibinfo{title}{On the variance of
  the adaptive learning rate and beyond}} in
  \emph{\bibinfo{booktitle}{International Conference on Learning
  Representations}}.

\bibitem[{Massoudi \emph{et~al.}(2015)Massoudi, Van~Wanrooij, Versnel, and
  Van~Opstal}]{massoudi2015spectrotemporal}
\bibinfo{author}{Massoudi, R.}, \bibinfo{author}{Van~Wanrooij, M.~M.},
  \bibinfo{author}{Versnel, H.},  and \bibinfo{author}{Van~Opstal, A.~J.}
  (\textbf{\bibinfo{year}{2015}}). \enquote{\bibinfo{title}{Spectrotemporal
  response properties of core auditory cortex neurons in awake monkey}}
  \bibinfo{journal}{PLoS One} \textbf{10}(2), \bibinfo{pages}{e0116118}.

\bibitem[{McCowan \emph{et~al.}(2005)McCowan, Carletta, Kraaij, Ashby, Bourban,
  Flynn, Guillemot, Hain, Kadlec, Karaiskos \emph{et~al.}}]{mccowan2005ami}
\bibinfo{author}{McCowan, I.}, \bibinfo{author}{Carletta, J.},
  \bibinfo{author}{Kraaij, W.}, \bibinfo{author}{Ashby, S.},
  \bibinfo{author}{Bourban, S.}, \bibinfo{author}{Flynn, M.},
  \bibinfo{author}{Guillemot, M.}, \bibinfo{author}{Hain, T.},
  \bibinfo{author}{Kadlec, J.}, \bibinfo{author}{Karaiskos, V.} \emph{et~al.}
  (\textbf{\bibinfo{year}{2005}}). \enquote{\bibinfo{title}{The ami meeting
  corpus}} in \emph{\bibinfo{booktitle}{Proceedings of Measuring Behavior 2005,
  5th International Conference on Methods and Techniques in Behavioral
  Research}}, \bibinfo{organization}{Noldus Information Technology}, pp.
  \bibinfo{pages}{137--140}.

\bibitem[{McCracken and Sheldon(1997)}]{mccracken1997avian}
\bibinfo{author}{McCracken, K.~G.},  and \bibinfo{author}{Sheldon, F.~H.}
  (\textbf{\bibinfo{year}{1997}}). \enquote{\bibinfo{title}{Avian vocalizations
  and phylogenetic signal}} \bibinfo{journal}{Proceedings of the National
  Academy of Sciences} \textbf{94}(8), \bibinfo{pages}{3833--3836}.

\bibitem[{McDermott(2018)}]{mcdermott2018audition}
\bibinfo{author}{McDermott, J.~H.} (\textbf{\bibinfo{year}{2018}}).
  \enquote{\bibinfo{title}{Audition}} \bibinfo{journal}{Stevens' Handbook of
  Experimental Psychology and Cognitive Neuroscience} \textbf{2},
  \bibinfo{pages}{1--57}.

\bibitem[{Mesgarani \emph{et~al.}(2006)Mesgarani, Slaney, and
  Shamma}]{mesgarani2006discrimination}
\bibinfo{author}{Mesgarani, N.}, \bibinfo{author}{Slaney, M.},  and
  \bibinfo{author}{Shamma, S.~A.} (\textbf{\bibinfo{year}{2006}}).
  \enquote{\bibinfo{title}{Discrimination of speech from nonspeech based on
  multiscale spectro-temporal modulations}} \bibinfo{journal}{IEEE Transactions
  on Audio, Speech, and Language Processing} \textbf{14}(3),
  \bibinfo{pages}{920--930}.

\bibitem[{Meyer \emph{et~al.}(2017)Meyer, Williamson, Linden, and
  Sahani}]{meyer2017models}
\bibinfo{author}{Meyer, A.~F.}, \bibinfo{author}{Williamson, R.~S.},
  \bibinfo{author}{Linden, J.~F.},  and \bibinfo{author}{Sahani, M.}
  (\textbf{\bibinfo{year}{2017}}). \enquote{\bibinfo{title}{Models of neuronal
  stimulus-response functions: elaboration, estimation, and evaluation}}
  \bibinfo{journal}{Frontiers in systems neuroscience} \textbf{10},
  \bibinfo{pages}{109}.

\bibitem[{M{\l}ynarski and McDermott(2018)}]{mlynarski2018learning}
\bibinfo{author}{M{\l}ynarski, W.},  and \bibinfo{author}{McDermott, J.~H.}
  (\textbf{\bibinfo{year}{2018}}). \enquote{\bibinfo{title}{Learning midlevel
  auditory codes from natural sound statistics}} \bibinfo{journal}{Neural
  computation} \textbf{30}(3), \bibinfo{pages}{631--669}.

\bibitem[{M{\l}ynarski and McDermott(2019)}]{mlynarski2019ecological}
\bibinfo{author}{M{\l}ynarski, W.},  and \bibinfo{author}{McDermott, J.~H.}
  (\textbf{\bibinfo{year}{2019}}). \enquote{\bibinfo{title}{Ecological origins
  of perceptual grouping principles in the auditory system}}
  \bibinfo{journal}{Proceedings of the National Academy of Sciences}
  \textbf{116}(50), \bibinfo{pages}{25355--25364}.

\bibitem[{Nagrani \emph{et~al.}(2017)Nagrani, Chung, and
  Zisserman}]{nagrani2017voxceleb}
\bibinfo{author}{Nagrani, A.}, \bibinfo{author}{Chung, J.~S.},  and
  \bibinfo{author}{Zisserman, A.} (\textbf{\bibinfo{year}{2017}}).
  \enquote{\bibinfo{title}{Voxceleb: a large-scale speaker identification
  dataset}} \bibinfo{journal}{Telephony} \textbf{3}, \bibinfo{pages}{33--039}.

\bibitem[{Ondel \emph{et~al.}(2019)Ondel, Li, Sell, and
  Hermansky}]{ondel2019deriving}
\bibinfo{author}{Ondel, L.}, \bibinfo{author}{Li, R.}, \bibinfo{author}{Sell,
  G.},  and \bibinfo{author}{Hermansky, H.} (\textbf{\bibinfo{year}{2019}}).
  \enquote{\bibinfo{title}{Deriving spectro-temporal properties of hearing from
  speech data}} in \emph{\bibinfo{booktitle}{ICASSP 2019-2019 IEEE
  International Conference on Acoustics, Speech and Signal Processing
  (ICASSP)}}, \bibinfo{organization}{IEEE}, pp. \bibinfo{pages}{411--415}.

\bibitem[{Peyr{\'e} \emph{et~al.}(2019)Peyr{\'e}, Cuturi
  \emph{et~al.}}]{peyre2019computational}
\bibinfo{author}{Peyr{\'e}, G.}, \bibinfo{author}{Cuturi, M.} \emph{et~al.}
  (\textbf{\bibinfo{year}{2019}}). \enquote{\bibinfo{title}{Computational
  optimal transport: With applications to data science}}
  \bibinfo{journal}{Foundations and Trends{\textregistered} in Machine
  Learning} \textbf{11}(5-6), \bibinfo{pages}{355--607}.

\bibitem[{Pillow and Sahani(2019)}]{pillow2019editorial}
\bibinfo{author}{Pillow, J.},  and \bibinfo{author}{Sahani, M.}
  (\textbf{\bibinfo{year}{2019}}). \plainquote{\bibinfo{title}{Editorial
  overview: Machine learning, big data, and neuroscience}} .

\bibitem[{Ravanelli and Bengio(2018)}]{ravanelli2018speaker}
\bibinfo{author}{Ravanelli, M.},  and \bibinfo{author}{Bengio, Y.}
  (\textbf{\bibinfo{year}{2018}}). \enquote{\bibinfo{title}{Speaker recognition
  from raw waveform with sincnet}} in \emph{\bibinfo{booktitle}{2018 IEEE
  Spoken Language Technology Workshop (SLT)}}, \bibinfo{organization}{IEEE},
  pp. \bibinfo{pages}{1021--1028}.

\bibitem[{Saddler \emph{et~al.}(2020)Saddler, Gonzalez, and
  McDermott}]{saddler2020deep}
\bibinfo{author}{Saddler, M.~R.}, \bibinfo{author}{Gonzalez, R.},  and
  \bibinfo{author}{McDermott, J.~H.} (\textbf{\bibinfo{year}{2020}}).
  \enquote{\bibinfo{title}{Deep neural network models reveal interplay of
  peripheral coding and stimulus statistics in pitch perception}}
  \bibinfo{journal}{bioRxiv} .

\bibitem[{Salamon and Bello(2017)}]{salamon2017deep}
\bibinfo{author}{Salamon, J.},  and \bibinfo{author}{Bello, J.~P.}
  (\textbf{\bibinfo{year}{2017}}). \enquote{\bibinfo{title}{Deep convolutional
  neural networks and data augmentation for environmental sound
  classification}} \bibinfo{journal}{IEEE Signal Processing Letters}
  \textbf{24}(3), \bibinfo{pages}{279--283}.

\bibitem[{Salamon \emph{et~al.}(2014)Salamon, Jacoby, and
  Bello}]{salamon2014dataset}
\bibinfo{author}{Salamon, J.}, \bibinfo{author}{Jacoby, C.},  and
  \bibinfo{author}{Bello, J.~P.} (\textbf{\bibinfo{year}{2014}}).
  \enquote{\bibinfo{title}{A dataset and taxonomy for urban sound research}} in
  \emph{\bibinfo{booktitle}{Proceedings of the 22nd ACM international
  conference on Multimedia}}, pp. \bibinfo{pages}{1041--1044}.

\bibitem[{Santoro \emph{et~al.}(2017)Santoro, Moerel, De~Martino, Valente,
  Ugurbil, Yacoub, and Formisano}]{santoro2017reconstructing}
\bibinfo{author}{Santoro, R.}, \bibinfo{author}{Moerel, M.},
  \bibinfo{author}{De~Martino, F.}, \bibinfo{author}{Valente, G.},
  \bibinfo{author}{Ugurbil, K.}, \bibinfo{author}{Yacoub, E.},  and
  \bibinfo{author}{Formisano, E.} (\textbf{\bibinfo{year}{2017}}).
  \enquote{\bibinfo{title}{Reconstructing the spectrotemporal modulations of
  real-life sounds from fmri response patterns}} \bibinfo{journal}{Proceedings
  of the National Academy of Sciences} \textbf{114}(18),
  \bibinfo{pages}{4799--4804}.

\bibitem[{Sch{\"a}dler \emph{et~al.}(2012)Sch{\"a}dler, Meyer, and
  Kollmeier}]{schadler2012spectro}
\bibinfo{author}{Sch{\"a}dler, M.~R.}, \bibinfo{author}{Meyer, B.~T.},  and
  \bibinfo{author}{Kollmeier, B.} (\textbf{\bibinfo{year}{2012}}).
  \enquote{\bibinfo{title}{Spectro-temporal modulation subspace-spanning filter
  bank features for robust automatic speech recognition}} \bibinfo{journal}{The
  Journal of the Acoustical Society of America} \textbf{131}(5),
  \bibinfo{pages}{4134--4151}.

\bibitem[{Shamma(1996)}]{shamma1996auditory}
\bibinfo{author}{Shamma, S.~A.} (\textbf{\bibinfo{year}{1996}}).
  \enquote{\bibinfo{title}{Auditory cortical representation of complex acoustic
  spectra as inferred from the ripple analysis method}}
  \bibinfo{journal}{Network: Computation in Neural Systems} \textbf{7}(3),
  \bibinfo{pages}{439--476}.

\bibitem[{Singh and Theunissen(2003)}]{singh2003modulation}
\bibinfo{author}{Singh, N.~C.},  and \bibinfo{author}{Theunissen, F.~E.}
  (\textbf{\bibinfo{year}{2003}}). \enquote{\bibinfo{title}{Modulation spectra
  of natural sounds and ethological theories of auditory processing}}
  \bibinfo{journal}{The Journal of the Acoustical Society of America}
  \textbf{114}(6), \bibinfo{pages}{3394--3411}.

\bibitem[{Snyder \emph{et~al.}(2015)Snyder, Chen, and Povey}]{snyder2015musan}
\bibinfo{author}{Snyder, D.}, \bibinfo{author}{Chen, G.},  and
  \bibinfo{author}{Povey, D.} (\textbf{\bibinfo{year}{2015}}).
  \enquote{\bibinfo{title}{Musan: A music, speech, and noise corpus}}
  \bibinfo{journal}{arXiv preprint arXiv:1510.08484} .

\bibitem[{Snyder \emph{et~al.}(2018)Snyder, Garcia-Romero, Sell, Povey, and
  Khudanpur}]{snyder2018x}
\bibinfo{author}{Snyder, D.}, \bibinfo{author}{Garcia-Romero, D.},
  \bibinfo{author}{Sell, G.}, \bibinfo{author}{Povey, D.},  and
  \bibinfo{author}{Khudanpur, S.} (\textbf{\bibinfo{year}{2018}}).
  \enquote{\bibinfo{title}{X-vectors: Robust dnn embeddings for speaker
  recognition}} in \emph{\bibinfo{booktitle}{2018 IEEE International Conference
  on Acoustics, Speech and Signal Processing (ICASSP)}},
  \bibinfo{organization}{IEEE}, pp. \bibinfo{pages}{5329--5333}.

\bibitem[{Stevens \emph{et~al.}(1937)Stevens, Volkmann, and
  Newman}]{stevens1937scale}
\bibinfo{author}{Stevens, S.~S.}, \bibinfo{author}{Volkmann, J.},  and
  \bibinfo{author}{Newman, E.~B.} (\textbf{\bibinfo{year}{1937}}).
  \enquote{\bibinfo{title}{A scale for the measurement of the psychological
  magnitude pitch}} \bibinfo{journal}{The Journal of the Acoustical Society of
  America} \textbf{8}(3), \bibinfo{pages}{185--190}.

\bibitem[{Tanno \emph{et~al.}(2019)Tanno, Arulkumaran, Alexander, Criminisi,
  and Nori}]{tanno2019adaptive}
\bibinfo{author}{Tanno, R.}, \bibinfo{author}{Arulkumaran, K.},
  \bibinfo{author}{Alexander, D.}, \bibinfo{author}{Criminisi, A.},  and
  \bibinfo{author}{Nori, A.} (\textbf{\bibinfo{year}{2019}}).
  \enquote{\bibinfo{title}{Adaptive neural trees}} in
  \emph{\bibinfo{booktitle}{International Conference on Machine Learning}},
  \bibinfo{organization}{PMLR}, pp. \bibinfo{pages}{6166--6175}.

\bibitem[{Thoret \emph{et~al.}(2020)Thoret, Andrillon, L{\'e}ger, and
  Pressnitzer}]{thoret2020probing}
\bibinfo{author}{Thoret, E.}, \bibinfo{author}{Andrillon, T.},
  \bibinfo{author}{L{\'e}ger, D.},  and \bibinfo{author}{Pressnitzer, D.}
  (\textbf{\bibinfo{year}{2020}}). \enquote{\bibinfo{title}{Probing
  machine-learning classifiers using noise, bubbles, and reverse correlation}}
  \bibinfo{journal}{BioRxiv} .

\bibitem[{Ulyanov \emph{et~al.}(2016)Ulyanov, Vedaldi, and
  Lempitsky}]{ulyanov2016instance}
\bibinfo{author}{Ulyanov, D.}, \bibinfo{author}{Vedaldi, A.},  and
  \bibinfo{author}{Lempitsky, V.} (\textbf{\bibinfo{year}{2016}}).
  \enquote{\bibinfo{title}{Instance normalization: The missing ingredient for
  fast stylization}} \bibinfo{journal}{arXiv preprint arXiv:1607.08022} .

\bibitem[{Vuong \emph{et~al.}(2020)Vuong, Xia, and Stern}]{Vuong2020}
\bibinfo{author}{Vuong, T.}, \bibinfo{author}{Xia, Y.},  and
  \bibinfo{author}{Stern, R.~M.} (\textbf{\bibinfo{year}{2020}}).
  \enquote{\bibinfo{title}{{Learnable Spectro-Temporal Receptive Fields for
  Robust Voice Type Discrimination}}} in \emph{\bibinfo{booktitle}{Proc.
  Interspeech 2020}}, pp. \bibinfo{pages}{1957--1961},
  \dodoi{10.21437/Interspeech.2020-1878}.

\bibitem[{Williamson \emph{et~al.}(2016)Williamson, Ahrens, Linden, and
  Sahani}]{williamson2016input}
\bibinfo{author}{Williamson, R.~S.}, \bibinfo{author}{Ahrens, M.~B.},
  \bibinfo{author}{Linden, J.~F.},  and \bibinfo{author}{Sahani, M.}
  (\textbf{\bibinfo{year}{2016}}). \enquote{\bibinfo{title}{Input-specific gain
  modulation by local sensory context shapes cortical and thalamic responses to
  complex sounds}} \bibinfo{journal}{Neuron} \textbf{91}(2),
  \bibinfo{pages}{467--481}.

\bibitem[{Woolley \emph{et~al.}(2005)Woolley, Fremouw, Hsu, and
  Theunissen}]{woolley2005tuning}
\bibinfo{author}{Woolley, S.~M.}, \bibinfo{author}{Fremouw, T.~E.},
  \bibinfo{author}{Hsu, A.},  and \bibinfo{author}{Theunissen, F.~E.}
  (\textbf{\bibinfo{year}{2005}}). \enquote{\bibinfo{title}{Tuning for
  spectro-temporal modulations as a mechanism for auditory discrimination of
  natural sounds}} \bibinfo{journal}{Nature neuroscience} \textbf{8}(10),
  \bibinfo{pages}{1371--1379}.

\bibitem[{Yarkoni and Westfall(2017)}]{yarkoni2017choosing}
\bibinfo{author}{Yarkoni, T.},  and \bibinfo{author}{Westfall, J.}
  (\textbf{\bibinfo{year}{2017}}). \enquote{\bibinfo{title}{Choosing prediction
  over explanation in psychology: Lessons from machine learning}}
  \bibinfo{journal}{Perspectives on Psychological Science} \textbf{12}(6),
  \bibinfo{pages}{1100--1122}.

\bibitem[{Zhang \emph{et~al.}(2019)Zhang, Lucas, Ba, and
  Hinton}]{zhang2019lookahead}
\bibinfo{author}{Zhang, M.}, \bibinfo{author}{Lucas, J.}, \bibinfo{author}{Ba,
  J.},  and \bibinfo{author}{Hinton, G.~E.} (\textbf{\bibinfo{year}{2019}}).
  \enquote{\bibinfo{title}{Lookahead optimizer: k steps forward, 1 step back}}
  in \emph{\bibinfo{booktitle}{Advances in Neural Information Processing
  Systems}}, \bibinfo{publisher}{Curran Associates, Inc.}, Vol. 32, pp.
  \bibinfo{pages}{9597--9608}.

\end{thebibliography}

\end{document}